\documentclass[a4paper,11pt]{article}
\usepackage{jinstpub} 
\usepackage{lineno}
\usepackage{siunitx}

\title{\boldmath Accelerator beam phase space tomography using machine learning to account for
variations in beamline components}

\author[a,c,1]{A. Wolski,\note{Corresponding author.}}
\author[a,c]{D. Botelho,}
\author[b,c]{D. Dunning,}
\author[b,c]{A.E. Pollard}
\affiliation[a]{Department of Physics, University of Liverpool,\\ Oxford Street, Liverpool L69 7ZE, UK}
\affiliation[b]{ASTeC, STFC Daresbury Laboratory,\\ Daresbury, WA4 4AD, UK}
\affiliation[c]{The Cockcroft Institute,\\ Sci-Tech Daresbury, Keckwick Lane, Daresbury, WA4 4AD, UK}

\emailAdd{a.wolski@liverpool.ac.uk}

\abstract{We describe a technique for reconstruction of the four-dimensional transverse phase
space of a beam in an accelerator beamline, taking into account the presence of unknown errors on
the strengths of magnets used in the data collection. Use of machine learning allows rapid
reconstruction of the phase-space distribution while at the same time providing estimates of the
magnet errors. The technique is demonstrated using experimental data from CLARA, an accelerator
test facility at Daresbury Laboratory.}

\keywords{Beam dynamics; Beam optics; Analysis and statistical methods; Data reduction methods}

\begin{document}
\maketitle
\flushbottom

\section{Introduction: machine learning for accelerator beam phase space tomography}
\label{sec:intro}


Many modern accelerators rely on the production of high-quality
particle beams to reach their performance specifications. Facilities such
as X-ray free-electron lasers, for example, have demanding requirements
for electron beam transverse and longitudinal emittance
\cite{PELLEGRINI200333, mcneil2010, BARLETTA201069, seddon2017}.
Commissioning, tuning and effective operation of many accelerators
require the capability to make rapid and reliable measurements of beam
parameters at different points along a beamline, often starting at the particle
source.  In the case of the transverse beam emittance, quadrupole scans
provide a standard measurement technique
\cite{mintyzimmermann2003, PhysRevSTAB.17.052801}:
the emittance is found from the dependence of the beam size (measured
using, for example, a wire scanner or imaging screen) on the strength of
an upstream quadrupole magnet.  The use of tomographic methods
\cite{mckee1995,yakimenko2003,stratakis2003,stratakis2007,xiang2009,rohrs2009,xing2018,ji2019}
allows the transverse phase-space distribution of the beam to be
reconstructed from the screen images, yielding not just the emittance
but also the optics functions describing how the distribution changes
along the beamline.  In addition, phase space tomography can provide
information on coupling between transverse planes
\cite{hock2013,PhysRevAccelBeams.23.032804}
and on the detailed charge distribution within bunches. With the use
of RF transverse deflecting structures, it is possible to make detailed
measurements of the longitudinal phase space \cite{alesini2006, marx2019}.
Combining quadrupole scans with a transverse deflecting cavity makes
it possible to determine the five or six-dimensional phase-space beam
distribution \cite{jastermerz2021, scheinker2022, jastermerz2022, jastermerz2023}.

Despite the fact that phase space tomography based on quadrupole
scans is a well-established technique, there remain several significant
challenges in its application, especially for low-emittance beams.  Data
collection and processing can be time-consuming, especially for
higher-dimensional phase space reconstruction.  Traditional tomography
algorithms can be prone to artefacts in the reconstruction, especially
when the range or number of projection angles is limited (as is often
the case in beam phase space tomography)
\cite{Frikel_2013, HOCK201136}: although there are techniques that
can be used to reduce the appearance of reconstruction artefacts
(see, for example, \cite{PhysRevE.104.045201}), ultimately the fidelity
of the reconstruction
depends on collecting screen images with a sufficient number of different
strengths of the upstream quadrupoles, setting a lower limit on
the time needed for data collection and analysis.  Machine learning
(ML) techniques have been shown to be of value for tomography
\cite{wang2020}, including for phase space tomography in particle
accelerators
\cite{PhysRevAccelBeams.25.122803,PhysRevAccelBeams.26.104601},
but the use of ML for this purpose is relatively new and there is significant
scope for improvement in the methods and tools available.

Reconstruction of the beam phase-space distribution in an accelerator
using quadrupole scan data is further complicated by the possible
presence of errors in accelerator components.  For example, errors in
magnet strengths, screen calibration errors, beam energy errors or the
presence of dispersion (e.g.~from magnet alignment errors) can all have
a significant impact on the results.  Errors in the field strength in a magnet
may arise from hysteresis, calibration errors in magnet manufacture,
power supply or control system errors, or short-circuits within the coils of
the magnet. The beam energy may vary during a measurement as a
result of fluctuations in RF power amplitude or phase in linacs or an RF gun.
Although significant efforts are generally made to minimise
errors in accelerator systems, even a relatively small accelerator 
facility can contain hundreds or thousands of components, and access to
an accelerator by technical personnel to confirm that components are 
working correctly will involve interrupting machine operation during the
inspection.  As a result, even quite large errors can sometimes be
present in components used in quadrupole scan measurements, and beam
diagnostics techniques that allow for (and provide information on)
various errors in accelerator components would be of significant value
in accelerator tuning and operation.

One approach to accounting for variations in accelerator conditions
(including component errors) in diagnostics measurements is to
use a neural network that can be tuned to adapt to changes in the
relevant accelerator systems.  `Tunable' neural networks have been used
in high-energy physics \cite{baldi2016} and in the reconstruction of
phase-space distributions in accelerators \cite{PhysRevLett.130.145001}.
Scheinker et al.~\cite{scheinker2021,scheinker2021a,PhysRevE.107.045302} 
have proposed the use of a low-dimensional
latent space in an encoder-decoder architecture: by tuning the latent
space so that the decoder output matches the observations, the `true'
encoding of the quantity or property to be measured can be obtained.
For example, the input to the encoder may be a set of projections from
a beam distribution in phase space onto a co-ordinate axis, with the
output from the decoder being the full phase space distribution.  Given
the phase space distribution, the projection onto a co-ordinate axis is
readily found, and the latent space is adjusted (from a starting point
determined by the input to the encoder) so that the computed projection
matches as closely as possible the observed projection.  A similar
approach has been used by Roussel et al.~\cite{PhysRevLett.130.145001,
arXiv.2209.04505} in the reconstruction of the four-dimensional
phase-space distribution in a wakefield accelerator.

In the work presented in the current paper, we adapt the latent space
tuning technique to identify quadrupole strength errors in beam phase
space tomography measurements.  Our approach differs from previous work 
in that rather than tuning the neural network (or latent space) so that
the output matches as closely as possible the observations, we train a
network using data from a simulation that includes known errors.  This
avoids the need for an optimization routine for tuning the neural
network, and potentially allows information on the accelerator
conditions to be obtained directly in addition to the primary intended 
measurement. Although the technique is generally applicable to a wide 
range of different diagnostics on various types of accelerator, in this
paper we illustrate its use for reconstructing the charge distribution
in phase space in CLARA, an accelerator test facility at Daresbury
Laboratory \cite{clara1, PhysRevAccelBeams.23.044801}.
Phase space measurements in CLARA using ML have
been reported in an earlier paper \cite{PhysRevAccelBeams.25.122803}: in
the present work, we show the results of a new analysis of some of the 
previous data using the new method, based on neural networks trained on 
data from simulations including machine errors.  At the core of the 
technique is the use of an autoencoder (trained on simulated screen
images in the absence of magnet errors) to construct a latent
space that encodes both the observed screen images and the phase space
distribution. The latent space is extended by appending the magnet
errors: the beam distribution in phase space upstream of the magnets is
still described by just the original latent space; but by appending the
quadrupole errors we construct an `extended latent space' that encodes
the screen images in the presence of the quadrupole errors. A second
neural network is then trained to take the observed screen images in the
presence of magnet errors as input, providing the extended latent space
as output.  By this means, given a set of screen images from simulations
or measurements in the presence of magnet errors, both the phase
space distribution and the magnet errors can be found.

Results from phase space tomography can generally be validated by 
reconstructing observed screen images from simulations using the
phase space distribution obtained from tomographic analysis. We find
that with the method described in this paper, individual screen
images are not reproduced with the same level of detail as in the
previous reported analysis of data from CLARA:
this is to be expected, since
representing the screen images (or, equivalently, the phase space
distribution) using a relatively small set of components in a latent 
space inevitably entails some loss of information.  But importantly,
we find that the new approach leads to a phase space distribution that
better fits the overall observed variation in beam size over the course
of the quadrupole scan. The results of the extended latent space
technique also indicate significant errors in the quadrupole strengths,
which may partly be explained by an error in the beam energy.

We should emphasise that the goal of the technique presented here is
not to achieve a highly-detailed reconstruction, providing accurate
information on fine structures in the phase space distribution.  In
many situations (e.g.~when commissioning or tuning an accelerator) it
is more important to be able to predict accurately the overall variation
in beam size as the beam propagates along a beam line, and how the beam
size responds to changes in quadrupole strengths.  Especially in the 
early stages of accelerator commissioning or after restarting from a
shut-down period, it is also important to be able to identify
possible errors in accelerator components. The aim of the current work
is to demonstrate a technique that has the potential to meet these
needs. Development of the technique was partly motivated by the later 
analysis of the data reported in \cite{PhysRevAccelBeams.25.122803} that 
suggested significant errors on the magnets, or on the beam energy (or 
both).  In particular, it was found that if it was assumed that there
were some errors on the magnets used in the quadrupole scan, then the
phase space distribution obtained from the tomography analysis led to
a better fit to the observed variations in beam size over the course of 
the scan. However, with the usual tomography algorithms it is difficult
to perform a fully self-consistent analysis in a systematic way, and to
determine possible errors with any confidence.  This is because 
tomographic analysis depends on the assumed strengths of the magnets
used for the scan, and computing the phase space distribution with
errors on the magnets involves repeating the full analysis for each
set of magnet errors. For tomography in two (or more) degrees of 
freedom, the time required makes it impractical to use optimisation
routines to fit errors on the quadrupoles.  Even if a neural network
was used for the phase space reconstruction, different sets of training
data would be needed for different magnet errors. Using existing
techniques, fitting both the phase space distribution and the magnet
errors in a systematic and fully consistent way would be an extremely
time-consuming process. In practical situations, the time taken for data
processing and analysis is a significant consideration, and given the
possibility of (perhaps significant) errors on the components used in a
diagnostics measurement, the extended latent space technique that we
describe in this paper could be of value in tuning and operating
an accelerator.

The paper is organised as follows.
In section \ref{sec:machinelearningprinciples} we describe the
principles on which the ML technique described in the current work is
based.  There are
a number of possible variations on the structure of the neural networks, 
and the way in which the latent space is constructed: in section
\ref{sec:machinelearningmodel} we describe one option in more detail,
including examples of the training data and results from
the neural networks using simulated test data. As an illustration of the
application of the technique to experimental data, in section
\ref{sec:experimentaldemonstration} we present results
from CLARA.  Finally, in section \ref{sec:conclusions} we discuss
the potential benefits and possible limitations of the technique,
and outline potential areas for further development.

\section{Tomography with extended latent space to account for errors
on accelerator components}
\label{sec:machinelearningprinciples}


Phase space tomography aims to construct the distribution of particles
in phase space at a chosen point in an accelerator beamline (the
`reconstruction point') from images collected on a diagnostics screen
as a downstream location (the `observation point').  Between the
reconstruction point and the observation point, a number of quadrupoles
are used to control the evolution of the beam distribution.  By choosing
appropriate strengths for the quadrupoles, it is possible to rotate the
distribution in phase space, changing the projection of the phase
space distribution onto co-ordinate space: a rotation through 90$^\circ$
(for example) interchanges co-ordinate and momentum variables.
Thus, observing the co-ordinate distribution provided by the screen
image for a range of rotation angles (corresponding to different
quadrupole strengths) allows reconstruction of the distribution as a
function of both co-ordinate and momentum.

\begin{figure}[htbp]
\centering
\includegraphics[width=.95\textwidth]{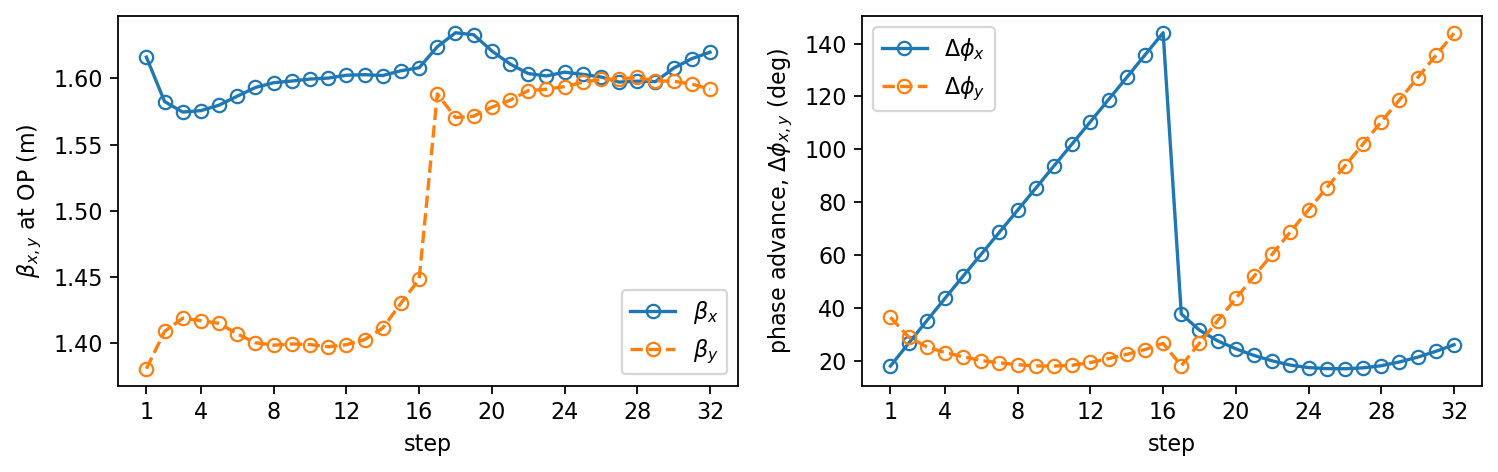}
\caption{Optics functions in a quadrupole scan on CLARA. Left:
horizontal (blue, solid line) and vertical (orange, dashed line) beta
functions at the observation point at each step in the quadrupole scan.
Right: horizontal and vertical phase advances from the reconstruction
point and the observation point. Ideally, the beta functions would be
equal and constant, and the phase advances would cover a full range of
180$^\circ$; however, the ideal values cannot be achieved in practice
because of limitations from the quadrupole strengths and spacings.
\label{fig:opticsfunctions}}
\end{figure}

As well as determining the rotation in phase space between the
reconstruction point and the observation point, the quadrupoles between
these two points will determine the horizontal and vertical beta
functions that characterise the size and shape of the image observed
on a diagnostics screen.  Generally, it is desirable to maintain an
aspect ratio close to 1 on the screen (i.e.~a roughly circular beam
image), and to avoid beam sizes that are either very small (approaching
the resolution limit of the imaging system) or very large (and as a
result, very diffuse and with low average intensity).  The quadrupole
strengths used in a scan for phase space tomography are therefore
determined by the need to cover a good range of rotation angles in
phase space while maintaining a reasonable size and shape for the image
on the diagnostic screen. For the work reported in this paper (including
experimental results from CLARA) the beta functions at the observation
point and the phase advances between the reconstruction point and the
observation point are shown in figure \ref{fig:opticsfunctions}.
Although there is some variation in the beta functions over the course
of the quadrupole scan, and the horizontal and vertical beta functions
are not exactly equal, there is not an absolute requirement for equal,
fixed beta functions, and if the beta functions at the reconstruction
point correctly describe the phase space distribution, then we expect
only small changes in the size and shape of the beam image at the
observation point. In practice, we observe much larger changes in size
and shape than expected (see, for example, the experimental results
shown later, in figure \ref{fig:sinogramautoencoderexample100pC}):
this indicates that the phase space
distribution differs from the design specifications.  One of the main
purposes of the tomography measurement is to determine the real
distribution. 

A further consideration concerns the sizes of the data structures used
for the sinograms (the set of images collected in a quadrupole scan) and
the phase space distribution.  This is a particular issue for
higher-dimensional reconstruction of the phase space distribution.  The
screen images could easily have a resolution of order 100$\times$100
pixels.  If the four-dimensional phase space is reconstructed to the
same resolution, the required data structure is a four-dimensional array
with a total of $10^8$ components.  While the processing of data sets of
this size (roughly 1\,Gb using double-precision arithmetic) is, in
principle, within the capability of modern computers, conventional
tomography algorithms will generally require a memory capacity several
orders of magnitude larger than the capacity needed just to store the result.
Even if machine learning is used, training a neural network with data
sets of this size places significant demands on computational power.
Image compression techniques
can be used to reduce the sizes of the data sets significantly; but
it is not clear how conventional tomography algorithms can be used with
images in compressed form.  However, it was shown in
\cite{PhysRevAccelBeams.25.122803} that ML tools can be used for
tomography with sinograms (sets of screen images) and phase space
distributions in compressed form: the only requirement is that the
same information is contained in the compressed image as in the original
image. Once the final result (the phase space distribution) is obtained,
then the data can be decompressed for visualisation and further 
analysis. Data compression can conveniently be achieved using discrete
cosine transforms (DCTs)
\cite{ahmednatarajanrao1974, chenpratt1984, raoyip1990}: using DCTs
it is possible to reduce the resolution by a factor (typically) between
five and ten, while retaining much of the information in the original
image.  This means that the memory required to store a four-dimensional
phase space distribution can be reduced by a factor $10^4$.
For the work presented in this paper, we use a DCT resolution of 16:
it is found that for the experimental measurements on CLARA, this
preserves a reasonable level of detail in the screen images and the
phase space distribution \cite{PhysRevAccelBeams.25.122803}.
In addition to compressing the images, we work in normalised phase
space: this simply involves scaling each screen image horizontally and
vertically by $1/\sqrt{\beta_x}$ and $1/\sqrt{\beta_y}$ respectively,
where $\beta_x$ and $\beta_y$ are the nominal horizontal and vertical
beta functions at the observation point, for the corresponding step
in the quadrupole scan (as shown in figure~\ref{fig:opticsfunctions}).
This simplifies the analysis, and can lead to improved accuracy in the
reconstructed phase space \cite{HOCK201136}.

\begin{figure}[htbp]
\centering
\includegraphics[width=.95\textwidth]{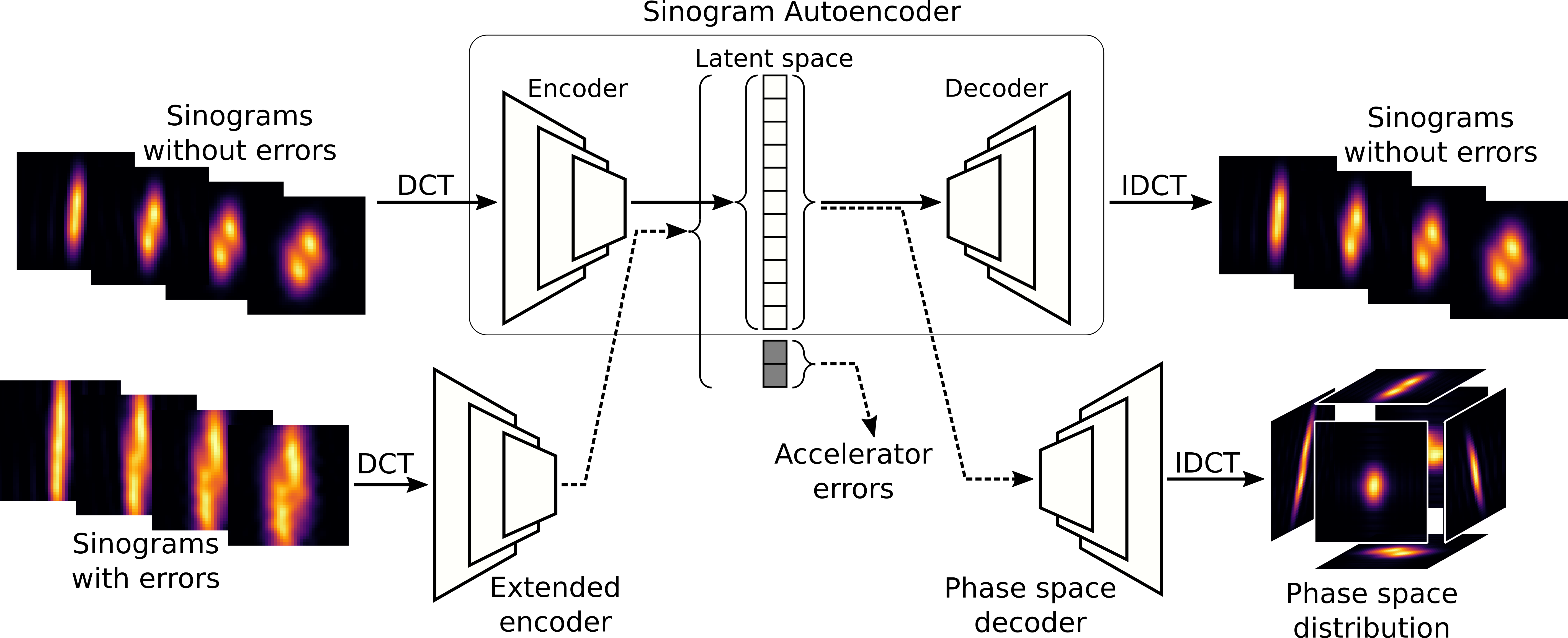} \\
(a) Latent space constructed using a sinogram autoencoder.\\[6mm]
\includegraphics[width=.95\textwidth]{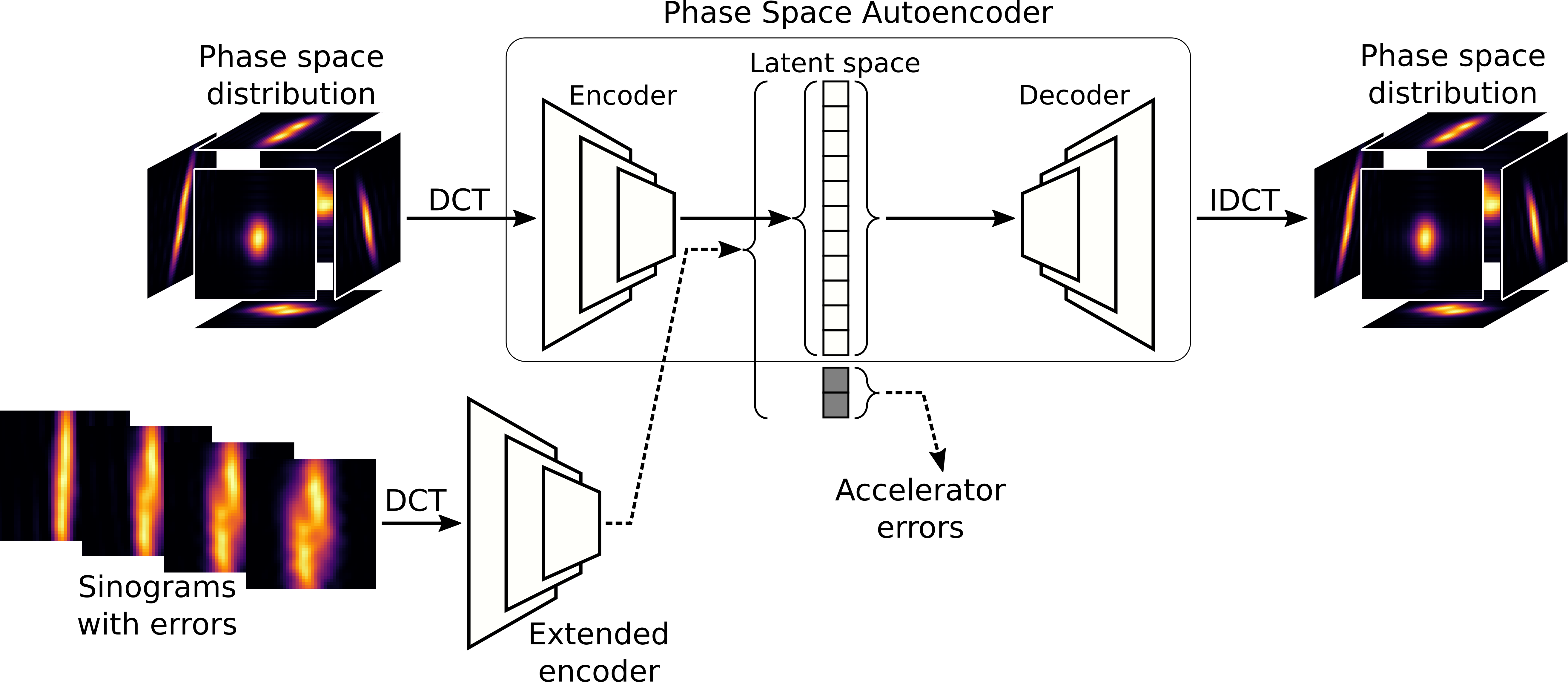} \\
(b) Latent space constructed using a phase space autoencoder.\\
\caption{Two possible structures for phase space tomography accounting
for errors on accelerator components, using ML methods. In each case,
sinograms consist of a sequence of beam images from a diagnostics
screen, collected over a scan of strengths of quadrupole magnets.
Data can be compressed using (for example) a discrete cosine
transform (DCT) and decompressed using an inverse DCT (IDCT).
In option (a) a sinogram autoencoder is trained on simulated data
without errors on accelerator components, and is used to construct the
``basic'' latent space (white boxes).  Corresponding sinograms with errors
on accelerator components are used to train an extended encoder, to produce an
extended latent space (white boxes plus gray boxes).  The phase space
decoder takes the basic latent space and produces the beam distribution
in four-dimensional phase space.  In option (b), the basic latent space is
constructed using an autoencoder taking the phase space distribution as input. 
\label{fig:machinelearningstrategy}}
\end{figure}

The technique we describe in this paper, for phase space tomography
taking into account errors on accelerator components, involves three 
related neural networks. The relationships between the networks are
illustrated schematically in
figure~\ref{fig:machinelearningstrategy}(a). There are many possible
variations on the technique: one example is shown in
figure~\ref{fig:machinelearningstrategy}(b); however, for simplicity,
we focus on the first option, with a latent space obtained using a
sinogram autoencoder.  In this case, the technique is applied as
follows. The first neural network is a `sinogram autoencoder': this 
takes a set of (DCT compressed) screen images as input, and
is trained to give the same screen images as output. Between the
input and output layers are a number of hidden layers.  The central
hidden layer consists of a relatively small number (some tens) of
nodes: this is the latent space, which encodes the main features of
the sinograms from the training data.  For the autoencoder, the
training data are constructed from a simulation of the quadrupole
scan in an accelerator tracking code, assuming that there are no
errors on any of the accelerator components.

The next step is to construct a second neural network, which we
refer to as the `extended encoder'.  This takes a sinogram as input,
and is trained to provide the extended latent space as output.
The training data in this case consists of sinograms produced in the
same way as for the training data for the sinogram autoencoder and with
the same initial phase space distributions, but now with random
errors on the quadrupole strengths.  The extended latent space is
constructed from the latent space from the autoencoder, with
additional values (nodes) corresponding to the quadrupole strength
errors.

Finally, we construct a third neural network, a `phase space
decoder'.  This takes the (unextended, or `basic') latent space
as input, and provides the (DCT compressed) four-dimensional
transverse phase space distribution as output.

Once the three neural networks have been trained, a sinogram
compiled from images recorded in a quadrupole scan on the
accelerator can be provided as input to the extended encoder.
The output consists of the latent space encoding the sinogram
that would be observed in the absence of accelerator component
errors (i.e.~the basic latent space), together with a set of values
representing the errors present on the components used in the
quadrupole scan.  The phase space distribution is obtained from the
phase space decoder, providing the basic latent space as input.
The results may be validated by simulating the quadrupole scan in
a tracking simulation, using the phase space distribution obtained
from the phase space decoder as the starting distribution, but now
including the errors on the accelerator components obtained from
the extended latent space. The accuracy of the reconstruction of
the phase space distribution may be assessed from the level of
agreement between the simulated screen images and the observed
images.

The important feature in this procedure is the use of an extended latent
space, constructed from two distinct sets of parameters.  The first
set of parameters encodes not only the images that would be observed
in a quadrupole scan in the absence of errors on accelerator components,
but also the beam distribution in phase space. The second set of
parameters in the extended latent space encodes the errors on the
accelerator components that were present during the collection of beam
images during a quadrupole scan. The basic principle of an extended
latent space can be implemented in various ways, with different relationships
between the neural networks. The particular method just described, and
discussed in further detail in the following sections, is illustrated in
figure~\ref{fig:machinelearningstrategy}(a); another possible
implementation is shown in figure~\ref{fig:machinelearningstrategy}(b).
In the latter case, the basic latent space is constructed from an 
autoencoder for the phase space distribution. This avoids the need for
a separate phase space decoder; however, we have found that an
autoencoder for the sinograms performs better for test data than an
autoencoder for the phase space distributions. This may be because of
differences in the sizes of the data sets, which has a direct impact on
the numbers of nodes needed in the layers of the neural network.

In a further variation, the basic latent space could be constructed not
from an autoencoder, but from principal component analysis of the
sinograms (without errors on the accelerator components) or of the
phase space distributions: autoencoders were originally proposed as
an alternative to principal component analysis for
dimensionality reduction \cite{https://doi.org/10.1002/aic.690370209}.
We have found that constructing the latent space using an autoencoder
usually leads to a reconstruction of test data that is more accurate
than if principle component analysis is used, though the difference
is not very great. For phase space tomography
on CLARA, the procedure represented in
figure~\ref{fig:machinelearningstrategy}(a) appears to work somewhat
better than alternatives we have tried, but it is possible that in other 
cases (for example, with differences in the design of the experiment, or
for different machine conditions) an alternative technique may produce
better results.

Application of the extended latent space technique in a real machine
relies on the validity of a number of assumptions.  One important
assumption is that the basic latent space provides an accurate
description of both the sinograms (without errors on accelerator
components) and the corresponding phase space
distributions.  This requires the quadrupole scan to cover a wide range
of phase space rotation angles (ideally, 180$^\circ$ in each degree of
freedom), with a sufficient number of steps.  If the range of angles is
too small or if there are two few steps, then the phase space
distribution will not be fully characterised by the sinogram, and the
problem of finding a distribution that reproduces a given sinogram will
be underconstrained.

\begin{figure}[htbp]
\centering
\includegraphics[width=.95\textwidth]{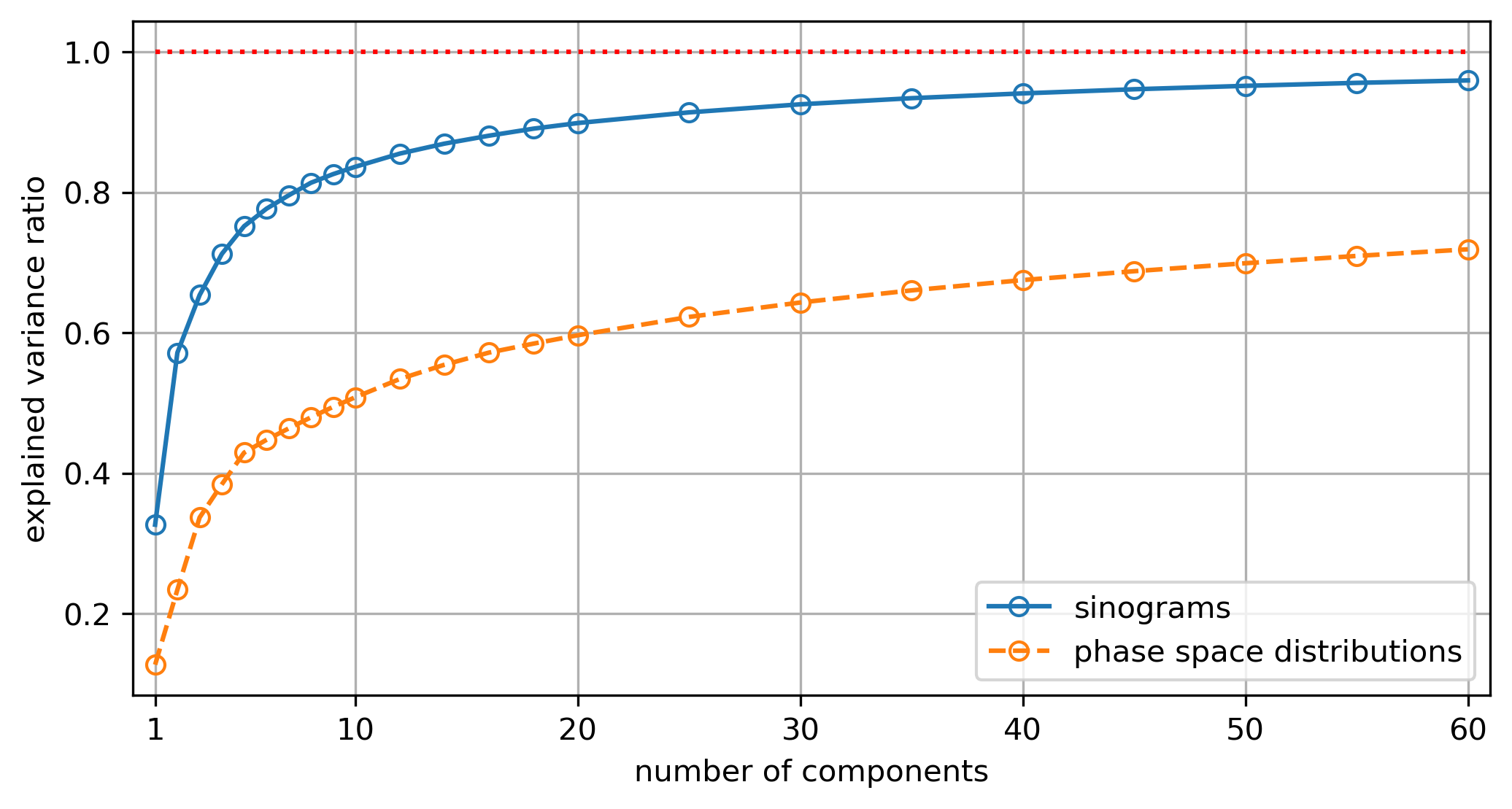}
\caption{Explained variance ratios from principal component analysis (PCA) of
the sinograms and phase space distributions in a set of training data. For a
given number of components in the PCA, the explained variance ratio for the
sinograms is higher than for the phase space distributions, indicating that the
sinograms can be represented more accurately by a set of principal components
than the phase space distributions. This may be a result of the fact that taking
a finite set of two-dimensional projections from  the four-dimensional phase
space inevitably results in the loss of some information about the phase space
distribution.
\label{fig:explainedvarianceratios}}
\end{figure}

An indication of the extent to which the phase space distribution is
determined by a sinogram can be obtained from principal component analysis
(PCA). In particular, we calculate the explained variance ratio for
different numbers of PCA components for the sinograms and the phase
space distributions: the results for the case considered in this paper
(with 32 steps in a quadrupole scan) are shown in
figure~\ref{fig:explainedvarianceratios}.  Using (for example) 40
PCA components, the explained variance ratio for the sinograms is 0.94,
compared to 0.68 for the phase space distributions.  This means that
using a latent space with 40 components, we would expect to be able to
describe the sinograms with significantly greater accuracy than the
phase space distributions.  To improve the accuracy of description of
the phase space distributions, we would need to increase the size of
the latent space.  However, the explained variance ratio increases
rather slowly when the number of PCA components is increased above 40,
and making the latent space too large is not desirable since this would
reduce the relative weight of the (small number of) errors on the
accelerator components when these are included by extending the latent
space. Alternatively, improving the accuracy of reconstruction of the
phase space distribution might be achieved by some modification of the
quadrupole scan, for example increasing the number of steps or
optimising the steps used. However, increasing the number of steps in
a quadrupole scan would increase the time needed for data collection on
the accelerator.  For the current work, we accept the limitations on the
accuracy of the reconstruction of the phase space suggested by the
explained variance ratios.

The extended latent space technique for phase space tomography also
assumes that the accelerator component errors included in the extended
latent space correspond to the errors affecting the screen images
observed in the accelerator.  When the training data for a neural
network are produced using a computer model of an accelerator, the
physical effects represented in the model are implicit in the training
data. This does not, of course, guarantee that results from the neural
network will always respect the physics in the model, although if the
network is designed and trained properly then the neural network results
should be close to the simulation results. But the fact that the training
data are constructed from simulations does mean that the neural network
cannot be expected to model physical behaviour not present in the
computational model of the accelerator. As an example, consider the case
that there is dispersion present in the section of beamline used in the
quadrupole scan when collecting data, but which is omitted from the model.
In the real accelerator, varying the quadrupole strengths will vary
the dispersion at the observation point, and because of the energy
spread on the beam, the variation in beam size in the accelerator
during the quadrupole scan will differ from the variation expected from
the computational model.  This will be the case even if all other aspects of 
the system are represented accurately. As a result, if the model is used to
reconstruct the phase space distribution from observed screen images, 
the distribution that is obtained will not exactly match the
distribution in the accelerator. Furthermore, using the reconstructed
phase space distribution in a simulation of the quadrupole scan will
produce screen images that differ from those observed experimentally.
The reconstructed phase space distribution can always be validated by
using the distribution in a simulation of the quadrupole scan, but
differences between the observed and simulated images may arise either
from limitations on the technique (for example, if the reconstruction of
the phase space distribution is not properly constrained by the
sinogram) or from differences between the physics in the real
accelerator and the physics included in the computational model.

\section{Implementation of the extended latent space technique: an example}
\label{sec:machinelearningmodel}


\subsection{Training data}
\label{sec:trainingdata}

\begin{figure}[htbp]
\centering
Case A \\
\includegraphics[width=.95\textwidth]{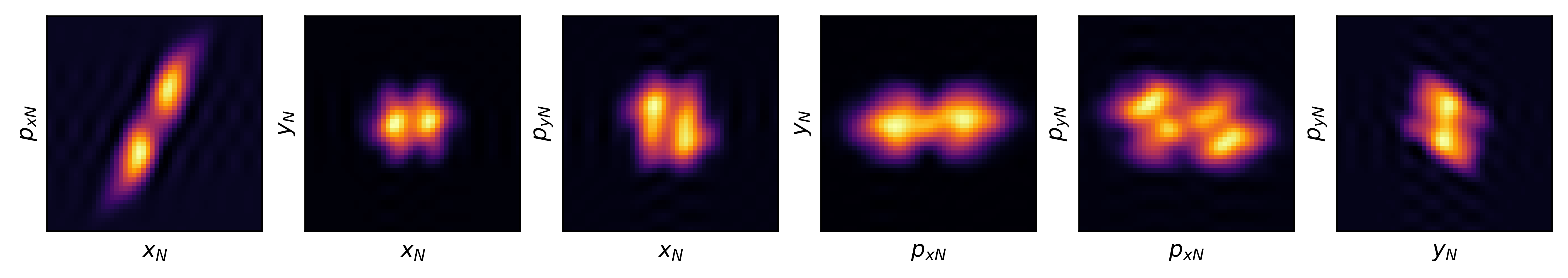} \\
Case B \\
\includegraphics[width=.95\textwidth]{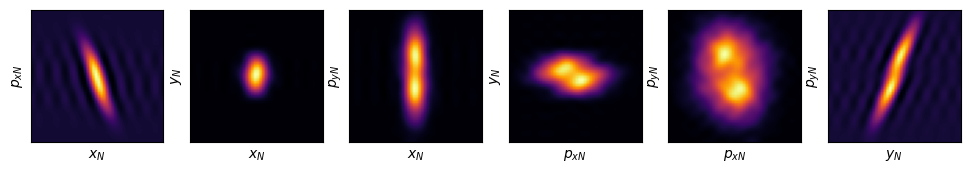} \\
Case C \\
\includegraphics[width=.95\textwidth]{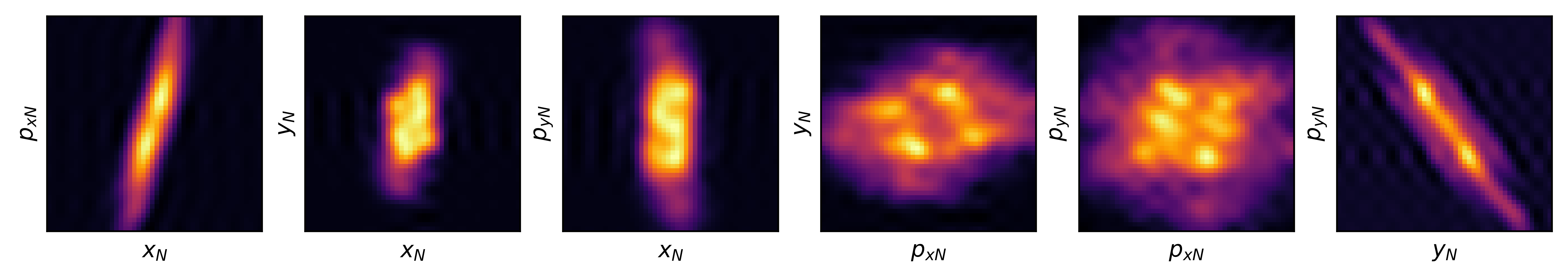} \\
\caption{Examples of phase space distributions from the training data,
constructed from Hermite--Gaussian modes using equations
(\ref{eq:hermitemode}) and (\ref{eq:hermitegaussiandistribution}).
Two-dimensional projections from the four-dimensional distribution onto
different pairs of axes in normalised phase space are shown.
\label{fig:phasespaceprojectionsexamples}}
\end{figure}

\paragraph{Phase space distributions}
The training data used for the neural networks should resemble
the phase space distribution that we expect to see in the
accelerator, while including sufficient variation to allow for some 
significant differences from the expected distribution.  Ideally, the
transverse phase space distribution in CLARA would be described by a 
four-dimensional Gaussian function.  In practice, we observe
some detailed structure in the screen images
\cite{PhysRevAccelBeams.25.122803}: the precise features depend on
the machine conditions and can also change over time.  The difficulty
in knowing in advance
what kinds of detailed structure may be observed presents some
challenges in generating appropriate training data. To allow for a
suitable range of possible distributions in phase space, we base the
training data on Hermite--Gaussian modes.  In normalised phase space
with co-ordinates $(x_N, p_{xN}, y_N, p_{yN})$, we first construct a 
Hermite mode $H_\mathbf{m}$:
\begin{equation}
H_\mathbf{m}(\mathbf{x}_N) = 
H_{m_x}\!\left( \frac{\sqrt{2}x_N}{w_\mathbf{m}} \right)
H_{m_{px}}\!\left( \frac{\sqrt{2}p_{xN}}{w_\mathbf{m}} \right)
H_{m_y}\!\left( \frac{\sqrt{2}y_N}{w_\mathbf{m}} \right)
H_{m_{py}}\!\left( \frac{\sqrt{2}p_{yN}}{w_\mathbf{m}} \right),
\label{eq:hermitemode}
\end{equation}
where $\mathbf{m}$ is a vector with components $(m_x, m_{px}, m_y, m_{py})$,
$H_m(\chi)$ is the Hermite polynomial of order $m$,
$\mathbf{x}_N$ is a vector in normalised phase space, and
$w_\mathbf{m}$ is a constant associated with the width of the 
distribution. The density of particles in phase space is obtained by 
summing over modes with random amplitudes $a_i$, and with random
four-dimensional rotations $\mathcal{R}_i$, `stretch' transformations
$\mathcal{S}_i$ and translations $\mathcal{T}_i$ applied to each mode:
\begin{equation}
\rho = \rho_0 \sum_\mathbf{i} a_i
\mathcal{R}_i \,
\mathcal{S}_i \,
\mathcal{T}_i \,
\left| H_{\mathbf{m}_i}(\mathbf{x}_N) \right|^2
 \exp\!\left( -\frac{\mathbf{x}_N^2}{w_{\mathbf{m}_i}^2} \right).
\label{eq:hermitegaussiandistribution}
\end{equation}
Here, $\rho_0$ is chosen so that the peak density is
$\mathrm{max}(\rho) = 1$.
For the training data used in the examples presented here, we use four
modes $\mathbf{m}_i$ for each case. Some examples of phase space
distributions from the training data constructed in this way are shown
in figure \ref{fig:phasespaceprojectionsexamples}.

\paragraph{Sinograms}
To construct the sinogram corresponding to a given phase space
distribution, a set of particles is created in the computer simulation
with the given distribution in phase space, and the co-ordinates of the
particles are transformed from the normalised phase space to the
accelerator phase space in the usual way:
\begin{equation}
\left(
\begin{array}{c}
x \\ p_x \\ y \\ p_y
\end{array}
\right) = 
\left(
\begin{array}{cccc}
\sqrt{\beta_x} & 0 & 0 & 0 \\
-\alpha_x / \sqrt{\beta_x} & 1/\sqrt{\beta_x} & 0 & 0 \\
0 & 0 & \sqrt{\beta_y} & 0 \\
0 & 0 & -\alpha_y / \sqrt{\beta_y} & 1/\sqrt{\beta_y}
\end{array}
\right)
\left(
\begin{array}{c}
x_N \\ p_{xN} \\ y_N \\ p_{yN}
\end{array}
\right).
\label{eq:normalisingtransformation}
\end{equation}
Here, $\beta_x$, $\alpha_x$, $\beta_y$ and $\alpha_y$ are the
Courant--Snyder parameters at the reconstruction point.  After
tracking particles from the reconstruction point to the observation
point, the co-ordinates of particles are transformed back to normalised
phase space, using the inverse of the transformation in
equation~(\ref{eq:normalisingtransformation}), with the appropriate
values for the Courant--Snyder parameters at the observation point.  A
simulated screen image is obtained by projecting the particle
distribution onto the $x$--$y$ plane.

\paragraph{Quadrupole errors}
For each phase space distribution, sinograms with and without errors
on the quadrupole magnets are constructed. Errors on the quadrupoles are
applied randomly, but are systematic in the sense that a given magnet
has a constant scaling factor applied to its strength at each step in
the quadrupole scan.  Thus, for a given case in the training data, the
strength of a magnet in the presence of errors is given by:
\begin{equation}
k_{1,\mathrm{err}} = (1 + e)k_1,
\end{equation}
where $k_1$ is the nominal quadrupole focusing strength,
$k_{1,\mathrm{err}}$ is the focusing strength in the presence of an
error, and $e$ is a random number from a flat distribution within a
chosen range.  For the case of CLARA, we choose a range
$-0.15 < e < 0.15$, i.e.~errors on the magnet strengths are in the
range $\pm 15\%$.  Although the magnet strengths in the accelerator are
expected to be within a much smaller range (in principle, less than $1\%$)
of their nominal values, a much larger range is used for the training data
partly to explore whether the the technique works with large errors, but
mainly to allow for the possibility of very large errors actually occuring in
practice.  Although a different error is applied to each of the five
quadrupoles in the scan, a non-zero mean error over the five magnets
could correspond to an error in the beam energy: the effect of an
increase in strength of all five magnets by 10\% would be similar
to the effect of a reduction in beam energy by 10\%. During experimental
data collection, nominal system settings were used for an intended
beam energy of \SI{35}{MeV}; however, because of time constraints the
actual beam energy was not confirmed by direct measurement.

\begin{figure}[htbp]
\centering
\includegraphics[width=.95\textwidth]{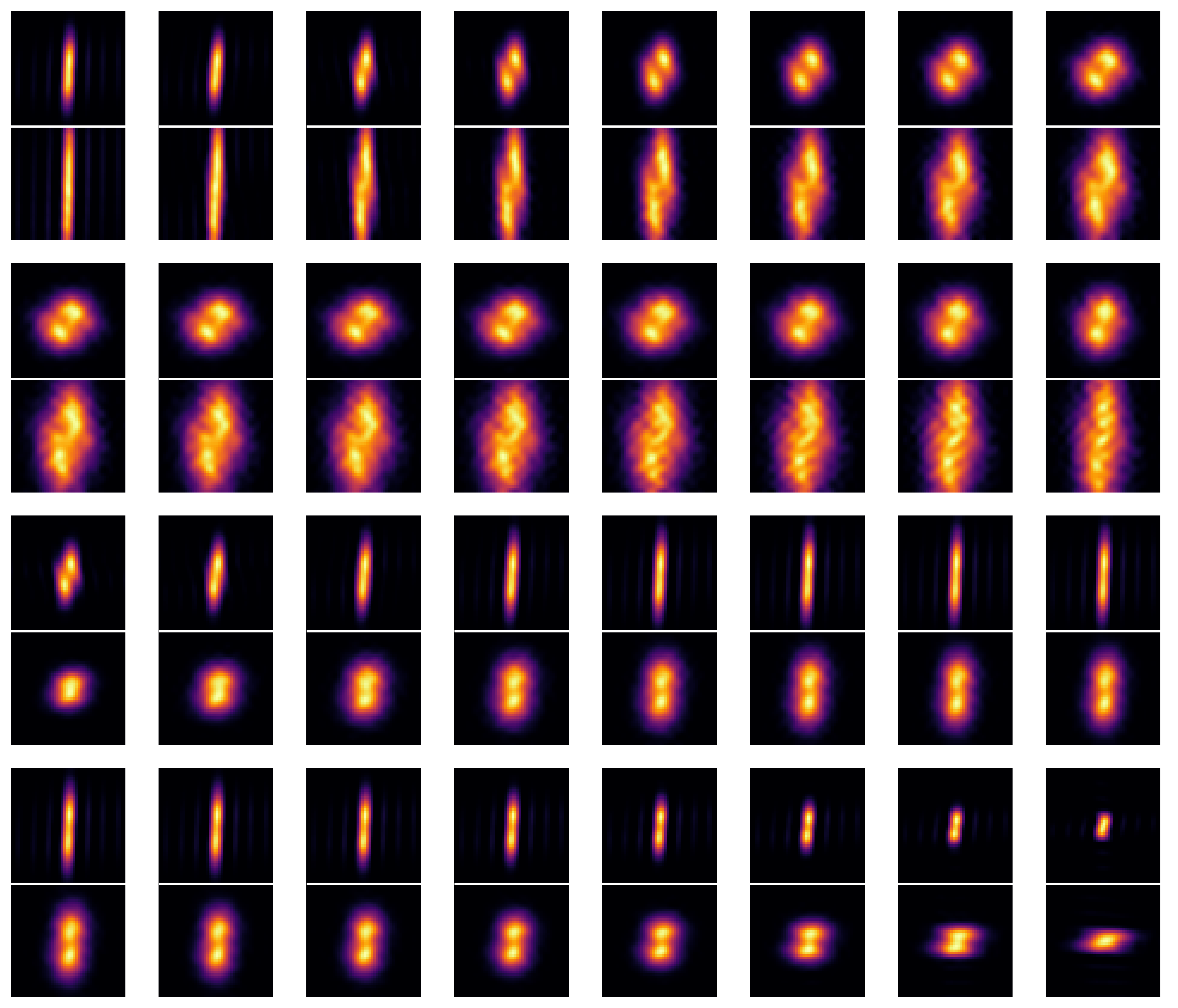}
\caption{Example of the impact of quadrupole errors on an observed
sinogram (Case B shown in figure
\ref{fig:phasespaceprojectionsexamples}).
The top image in each pair shows the beam image on a particular step
in the quadrupole scan, with zero errors on the quadrupole magnets; the
lower image in each pair shows the corresponding image with non-zero
errors on the quadrupole magnets. In this case, the errors on the
quadrupole magnets are -5.06\%, -12.8\%, 2.06\%, 4.17\% and -12.5\%.
\label{fig:sinogramquaderrorsexample}}
\end{figure}

Errors on the quadrupole magnets can have a significant impact on the
screen images at the observation point: this is illustrated (for simulated data)
in figure~\ref{fig:sinogramquaderrorsexample}. Note that the effect on the
beam images depends not just on the average error over all
quadrupoles, but also on how the errors vary from one quadrupole to the next.
For example, if two adjacent magnets have similar errors, then the effect
of the errors will depend on whether the magnets have the same polarity
(in which case the effects of the errors will combine) or opposite polarity
(in which case there may be some cancellation).

We note that the model used here for errors on the quadrupole magnets
is a greatly simplified representation of the errors that may occur in
practice.  For example, the error on a given magnet may vary with magnet
strength because of hysteresis effects: this may particularly be the
case when the polarity of a magnet changes sign.  Nonlinear or
time-dependent effects in the magnet power supplies may also lead to
complex behaviour of the magnet errors. To try to reduce hysteresis effects,
the order of steps in the quadrupole scan was designed to minimise the
number of times that any magnet would need to change polarity. However,
changes in polarity could not be avoided altogther and about half-way
through the scan, a change in polarity was needed for several of the
magnets: at this point, when collecting data experimentally, the magnet strengths
were cycled to return the cores of the magnets to a nominal operating
point on the magnetisation curve. The magnets were also cycled
immediately before a quadrupole scan was performed.

The computational model of the accelerator was used to construct
training data consisting of 20,000 phase space distributions together with the
corresponding sinograms.  A subset of 18,000 cases was
used in training the neural networks (the `training set'), and the
remaining 2,000 cases were reserved for testing the trained networks on
unseen data (the `test set').  The data were scaled so that each
sinogram (consisting of a set of 32 images) had DCT values in the
range -1 to +1, with the values representing the magnet errors lying
in the same range.

\subsection{Sinogram autoencoder}
\label{sec:sinogramautoencoder}

\begin{figure}[htbp]
\centering
\includegraphics[width=.95\textwidth]{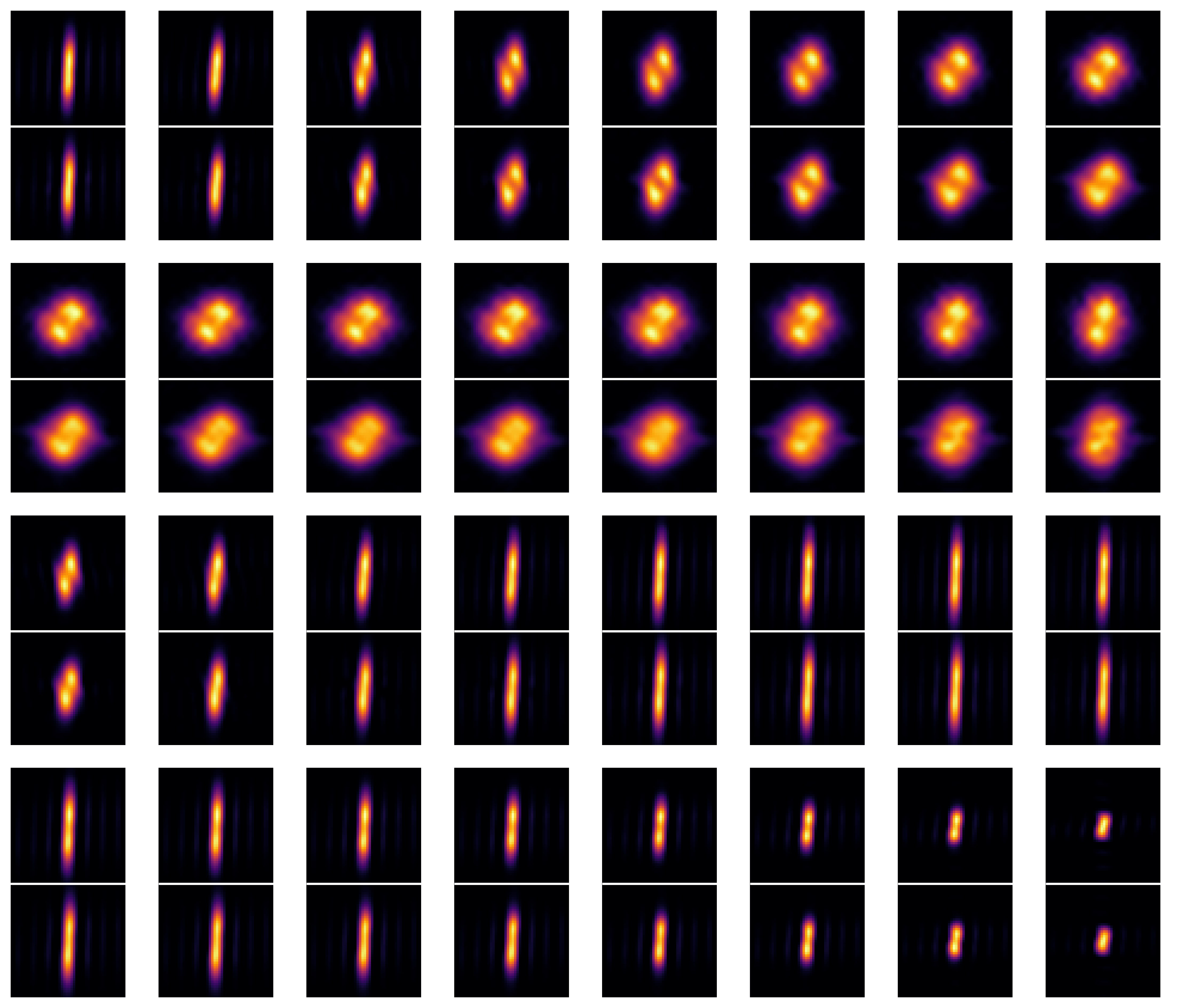}
\caption{Example of results from the sinogram autoencoder, for a case
from the set of test data (Case B shown in figure 
\ref{fig:phasespaceprojectionsexamples}).
In each pair, the top image shows the input image, constructed from
particle tracking; the lower image shows the output of the autoencoder.
\label{fig:sinogramautoencoderexample}}
\end{figure}

We use Keras \cite{chollet2015keras} for all the machine learning tasks
described in the current work. 
The first neural network to be constructed is the sinogram autoencoder.
The input and output data consist of the discrete cosine transform (DCT) 
of each simulated screen image, without quadrupole errors. The input and
output layers are flat layers consisting of 8,192 nodes, corresponding
to 32 images with DCT truncation to 16 components in each transverse
dimension. We use five hidden layers (with 1,600, 400, 40, 400 and
1,600 nodes): the latent space is given by the central layer, and so has
dimension 40. An example of results from the autoencoder with test data
is shown in figure \ref{fig:sinogramautoencoderexample}.

\begin{figure}[htbp]
\centering
\includegraphics[width=.85\textwidth]{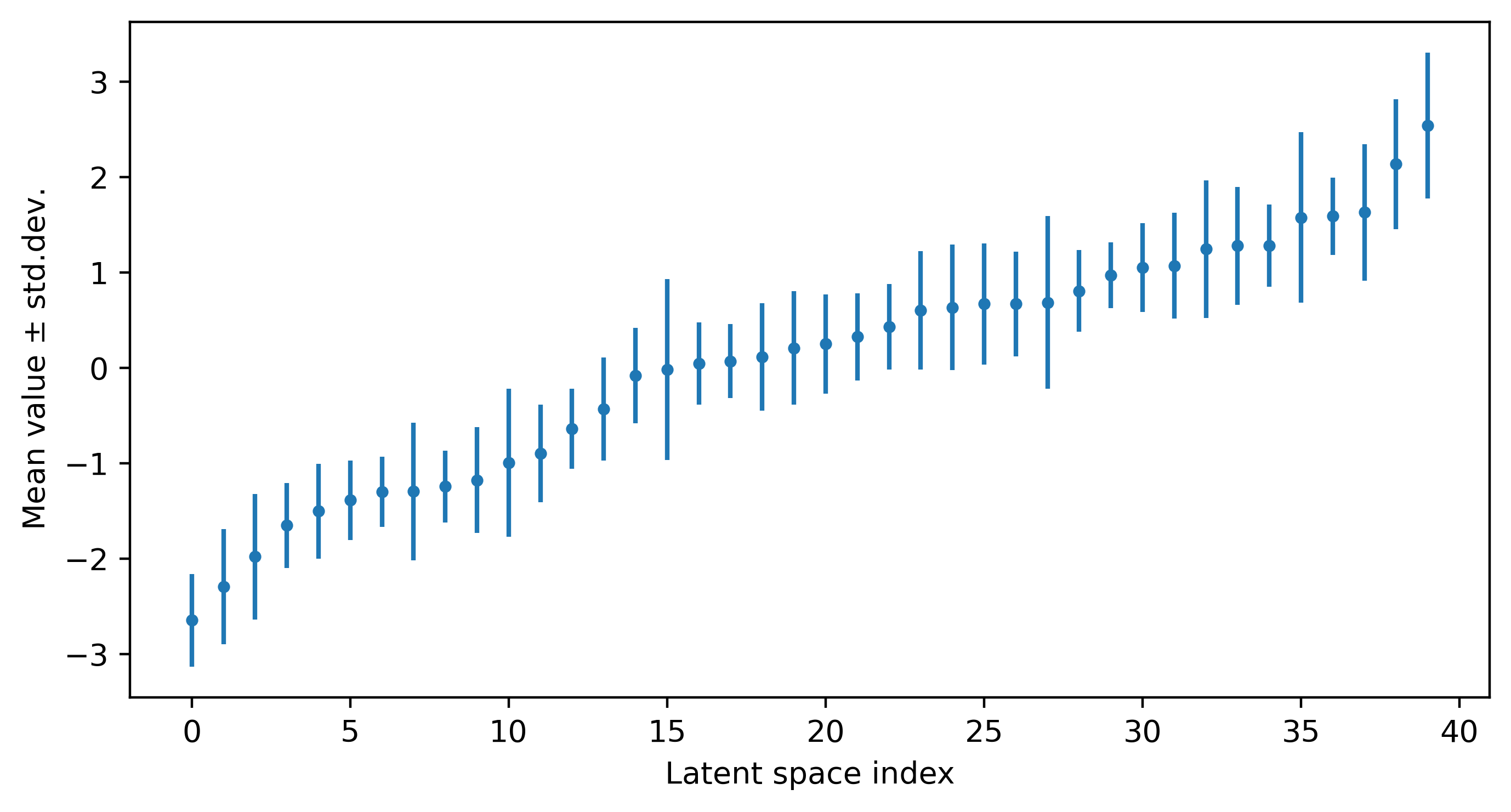}
\caption{Latent space values obtained from the sinogram autoencoder
for the training data.  The points show the mean values across the full
set of 20,000 cases (including training and test data); the vertical bars
show the standard deviation of each latent space parameter. For clarity,
latent space components are sorted in order of ascending mean value.
\label{fig:latentspace}}
\end{figure}

Using a latent space with 40 components, it is possible to describe even
quite detailed features in the sinograms with good accuracy: this is
consistent with the PCA results, which suggest an explained variance
ratio of 94\% with 40 PCA components.  The latent space values obtained
from the training data are shown in figure~\ref{fig:latentspace}.  Note
that for this plot, the latent space components have been sorted in
ascending order of the mean value across all cases in the training data.
The vertical bars show the standard deviation of each component.

\subsection{Extended encoder}
\label{sec:extendedencoder}

The next step is to construct an extended encoder to predict the
extended latent space (i.e.~including errors on the quadrupole magnets)
from a given sinogram.  The structure for the extended encoder is
similar to that of the first half of the sinogram autoencoder, and
consists of a flat input layer with 8,192 nodes, a single hidden layer,
and an output layer.  The output layer matches the extended latent
space: for the examples shown here, it consists of 40 nodes from the
latent space of the sinogram autoencoder, plus three nodes for the
quadrupole errors.  Note that although five quadrupole magnets are used
in the quadrupole scan, and errors are applied to all five magnets when
constructing the training data in simulation, the errors on the final 
two quadrupoles (which are close to the observation point) have little
impact on the images on the diagnostics screen.  As a result, although
it is possible to construct a neural network that can predict the
errors on all five magnets to good accuracy for the training data,
the accuracy of prediction for test data is generally very low.  We
therefore exclude these two magnets from the extended latent space, and
just use an extended latent space of 43 nodes.

For training the extended encoder, we double the size of the training
data set by combining the data sets with and without quadrupole errors:
the sinograms without errors can be considered a special case of the
sinograms with errors, in which the errors are zero.

For the extended encoder, we find that overfitting can be a problem:
training the network can result in a network that `predicts' the
extended latent space with very high accuracy for the data used in
training, but with relatively poor accuracy for the test data.
Reducing the size of the network helps to some extent: this is the
reason for using just a single hidden layer in the extended encoder. 
Other strategies for avoiding overfitting (such as the use of dropout
layers) appear to have little benefit; however, applying L2 regularisation
\cite{Krogh1991ASW, kerasL2} to the hidden layer appears to be effective,
and is used for the results presented here.

\begin{figure}[htbp]
\centering
\includegraphics[width=.95\textwidth]{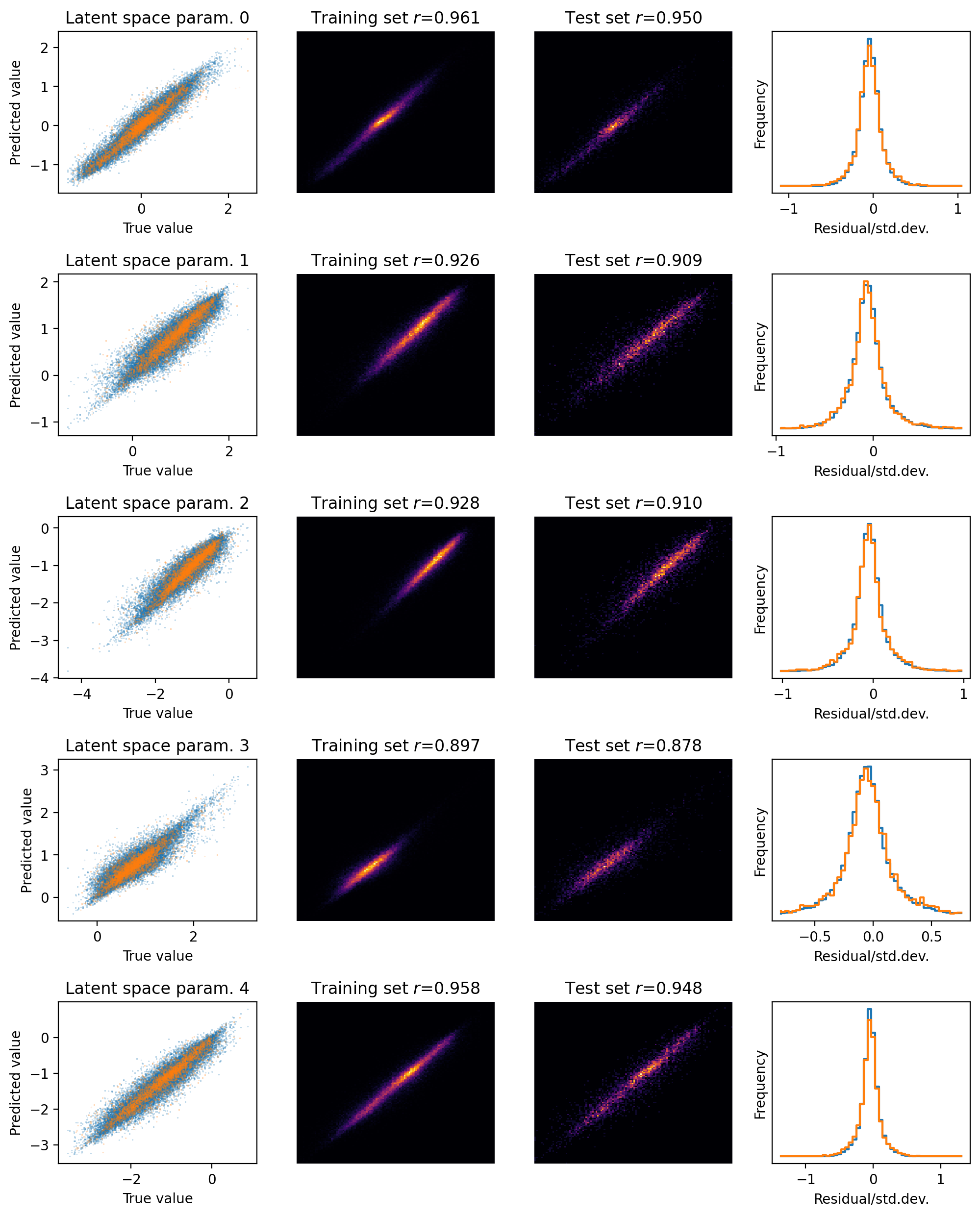}
\caption{Correlations between ground truth extended latent space values and values
predicted by the extended encoder, for the first five latent space parameters.  The left-hand
column shows the correlation for training data (blue points) and test data (orange points).
The second and third columns show the training data and test data separately, with
correlation coefficients ($r$ values).  The fourth column shows the distribution of
residuals (scaled by the standard deviation), with the blue line for the training
data, and the orange line for the test data.
\label{fig:sinogramencodercorrelations}}
\end{figure}

Correlations between the latent space values obtained from the sinogram
encoder and the values predicted by the extended encoder are shown in
figure~\ref{fig:sinogramencodercorrelations} for the first five
components of the latent space.  For all components, we find a good
correlation: for any given component of the latent space, the residuals
are small compared to the range of values (across all the training data)
for that component.  The residuals for the training set and the test set
from the training data show similar distributions.

\begin{figure}[htbp]
\centering
\includegraphics[width=.95\textwidth]{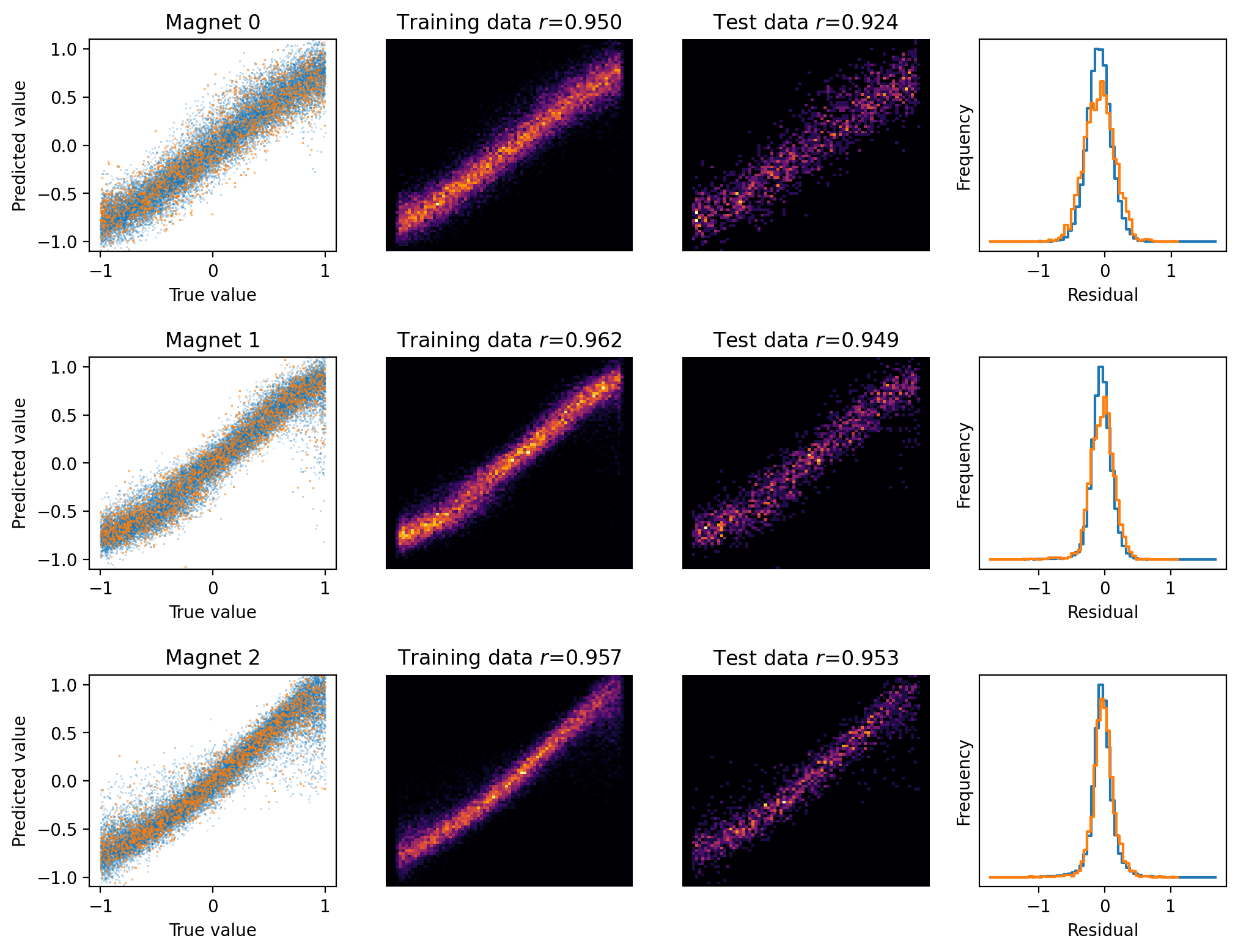}
\caption{Correlations between ground truth magnet errors and values for the errors
predicted by the extended encoder. The left-hand column shows the correlation for
training data (blue points) and test data (orange points).  The second and third columns
show the training data and test data separately, with correlation coefficients ($r$ values).
The fourth column shows the distribution of residuals, with the blue line for the training
data, and the orange line for the test data.
\label{fig:sinogramencodermagnetcorrelations}}
\end{figure}

Figure~\ref{fig:sinogramencodermagnetcorrelations} shows the correlations between the magnet errors and the values for the errors
predicted by the extended encoder.  Although there is
still a close correlation between the true values and the predicted
values, the residuals are larger than for the components of the extended
latent space describing the sinograms.  Uncertainties on values for
fitted magnet errors may be estimated from the standard deviations of
the residuals between ground truth values and predicted values. For the
test data, with scaled magnet errors in the range $\pm$1, the residuals
have standard deviations 0.225, 0.183 and 0.175 for magnets 0, 1 and 2
respectively. Since the errors are in the range $\pm$15\%, the absolute 
uncertainties on values for the magnet errors obtained from the extended
encoder are 3.4\%, 2.8\% and 2.6\% for magnets 0, 1 and 2 respectively.

\begin{figure}[htbp]
\centering
\includegraphics[width=.85\textwidth]{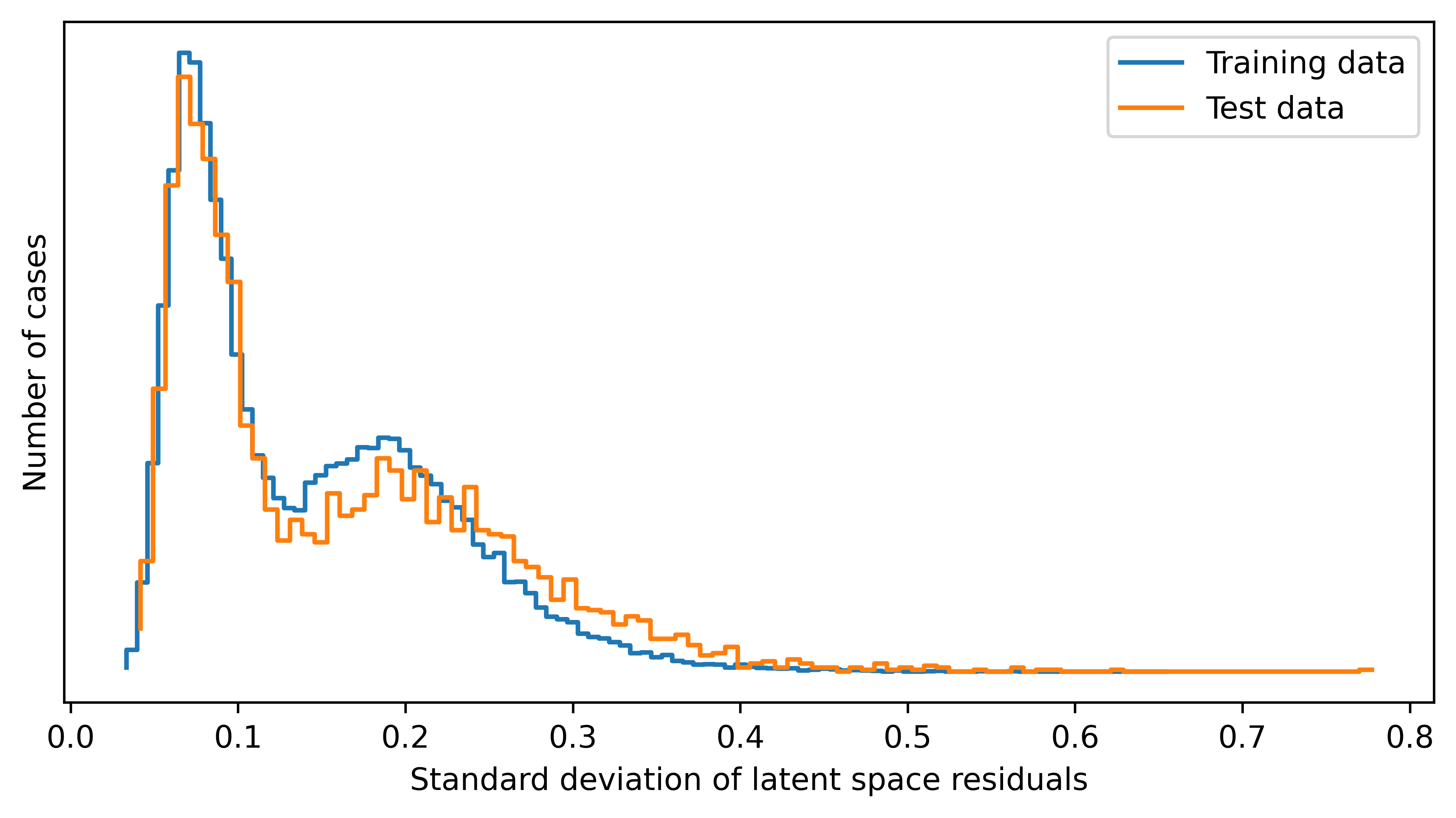}
\caption{Distribution (over all cases in the training data) of the standard deviations of
the residuals to the latent space from the sinogram encoder.  The slightly larger
residuals in the case of the test set, compared to the training set, suggests that there
may be some overfitting in the training of the sinogram encoder.  The two distinct peaks
correspond to cases in the training data with zero quadrupole errors (left-hand peak)
and with non-zero quadrupole errors (right-hand peak, with larger standard deviations).
\label{fig:sinogramencoderresidualsdistributions}}
\end{figure}

It is worth noting that the residuals for the predictions from the
extended encoder are larger for cases with magnet errors than for
cases without magnet errors.  This can be seen in
figure~\ref{fig:sinogramencoderresidualsdistributions}, which shows the
distribution of the standard deviation (taken over the 43 components of
the extended latent space) of the residuals of the extended encoder
predictions. The training set and test set are again shown separately,
so that it can be seen that the distributions for the two sets are
similar: the fact that the residuals for the test data tend to be a
little larger than the residuals for the training data suggests that
there may be some overfitting, but any overfitting appears to be
limited.  Both distributions show two distinct peaks, one below a
standard deviation residual 0.1, and another (broader) peak around 0.2.
Closer inspection of the data shows that the sharp peak below 0.1
corresponds to the cases in training data with zero quadrupole errors,
while the broader peak at a higher residual value corresponds to the
training data with non-zero quadrupole errors.

\subsection{Phase space decoder}
\label{sec:phasespacedecoder}

The final step is to construct the phase space decoder, a neural network
that takes as input the components of the extended latent space describing 
the sinograms (without quadrupole errors), and provides as output the 
four-dimensional transverse phase space distribution.  The input layer 
has 40 nodes, corresponding to the latent space of the sinogram autoencoder;
the output layer has $16^4 = 65,536$ nodes, corresponding 
to a DCT resolution of 16 in each dimension of the four-dimensional transverse
phase space. Between the input and 
output layers, there are three hidden layers (with 800, 1,200 and 2,400 
nodes).

\begin{figure}[htbp]
\centering
\includegraphics[width=\textwidth]{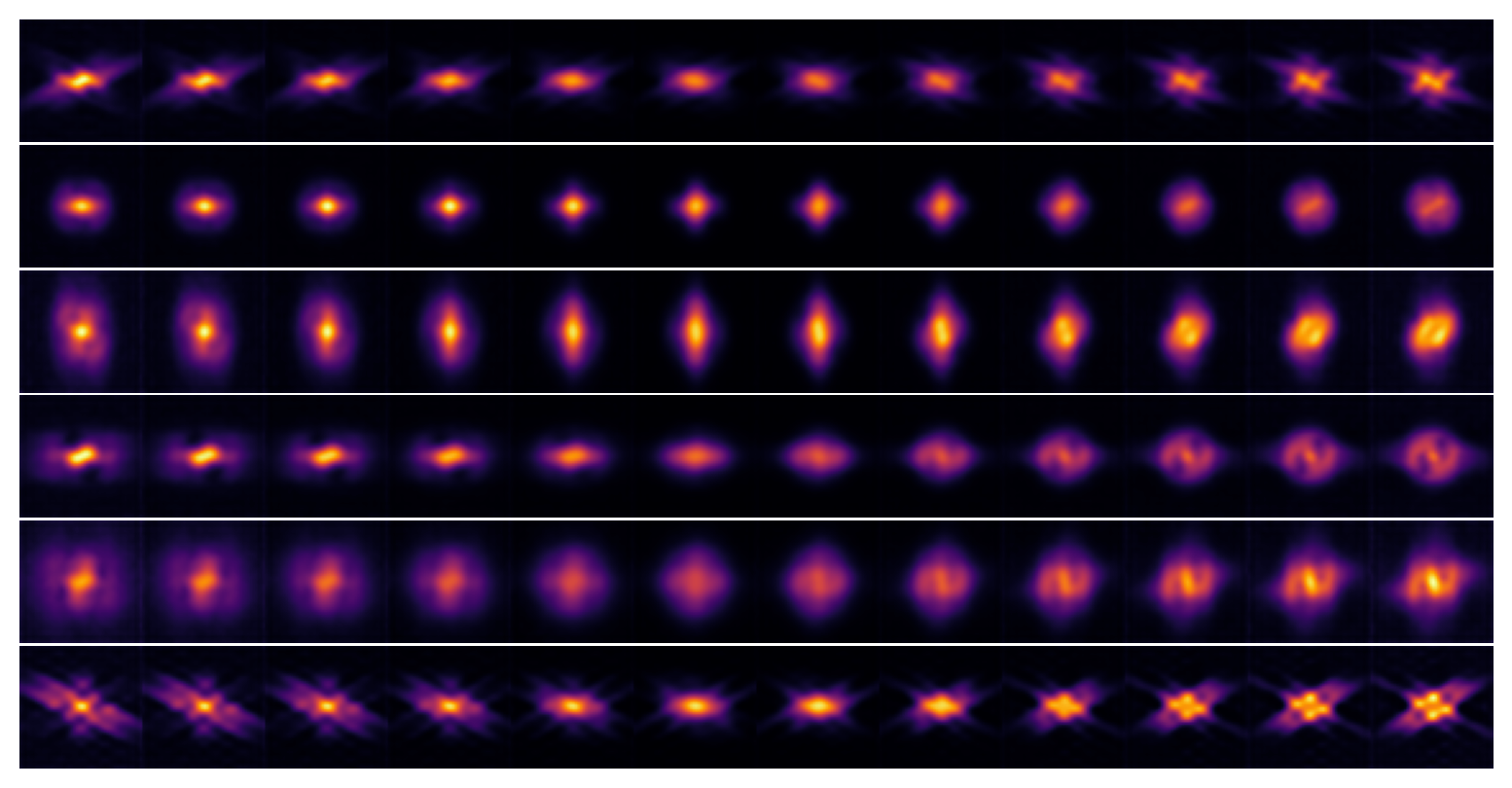}
\caption{Variation in phase space projections with variation in the
first latent space parameter. Each row shows the changes in a projection
of the phase space distribution as the first latent space parameter is
varied from $-5\sigma$ to $+5\sigma$ around its mean value (where
$\sigma$ is the standard deviation of the parameter over the set of
training data). All other latent space parameters are fixed at their
mean values. From top to bottom, the rows show projections onto
$p_x$ versus $x$, $y$ versus $x$, $p_y$ versus $x$, $y$ versus $p_x$,
$p_y$ versus $p_x$, and $p_y$ versus $y$. Inspection of similar plots
for other latent space parameters suggests that smooth variations in the
latent space parameters leads to smooth variations in the phase space
distribution.
\label{fig:latentspacevariation}}
\end{figure}

In using the latent space from the sinogram autoencoder to reconstruct
the phase space distribution, it is implicitly assumed that smooth
variations in the latent space parameters lead to smooth variations in
the phase space distribution. If small changes in the latent space
lead to large or abrupt changes in the phase space distribution, then
it will be difficult to reconstruct the phase space distribution with
any accuracy or reliability.  The way in which we have constructed the
sinogram autoencoder and phase space decoder does not guarantee that the
phase space distribution has a smooth dependence on the latent space;
however, it appears by inspection that the phase space distribution does
in fact vary smoothly in response to continuous changes in the latent
space parameters.  An example, showing how two-dimensional projections
of the four-dimensional phase space distribution vary in response to
changes in the first latent space parameter, is shown in
figure~\ref{fig:latentspacevariation}.

\begin{figure}[htbp]
\centering
Case A: quadrupole errors = (10.7\%, -5.01\%, -2.53\%, 7.50\%, 0.183\%) \\
\includegraphics[width=.77\textwidth]{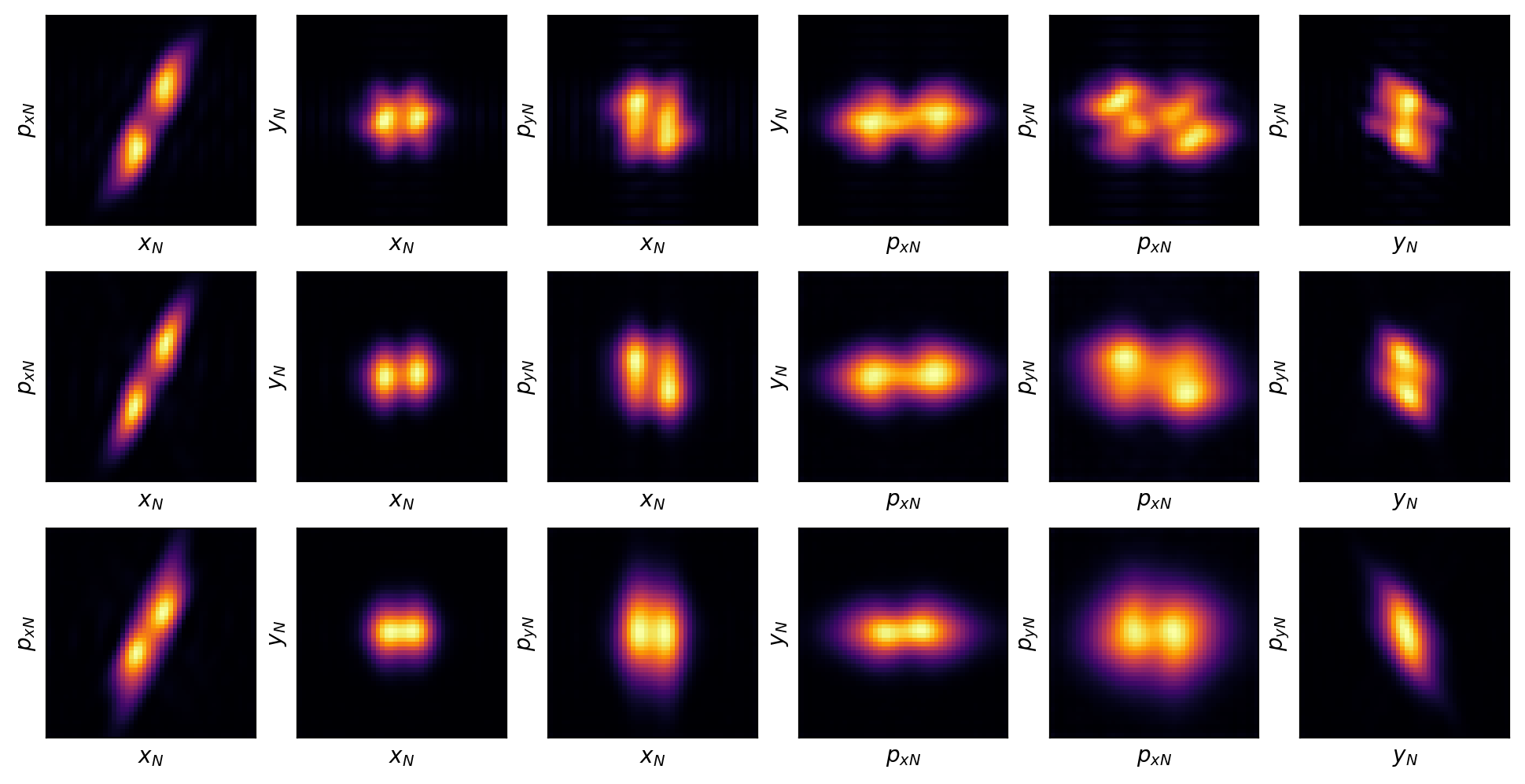} \\
Case B: quadrupole errors = (-5.06\%, -12.8\%, 2.06\%, 4.17\%, -12.5\%) \\
\includegraphics[width=.77\textwidth]{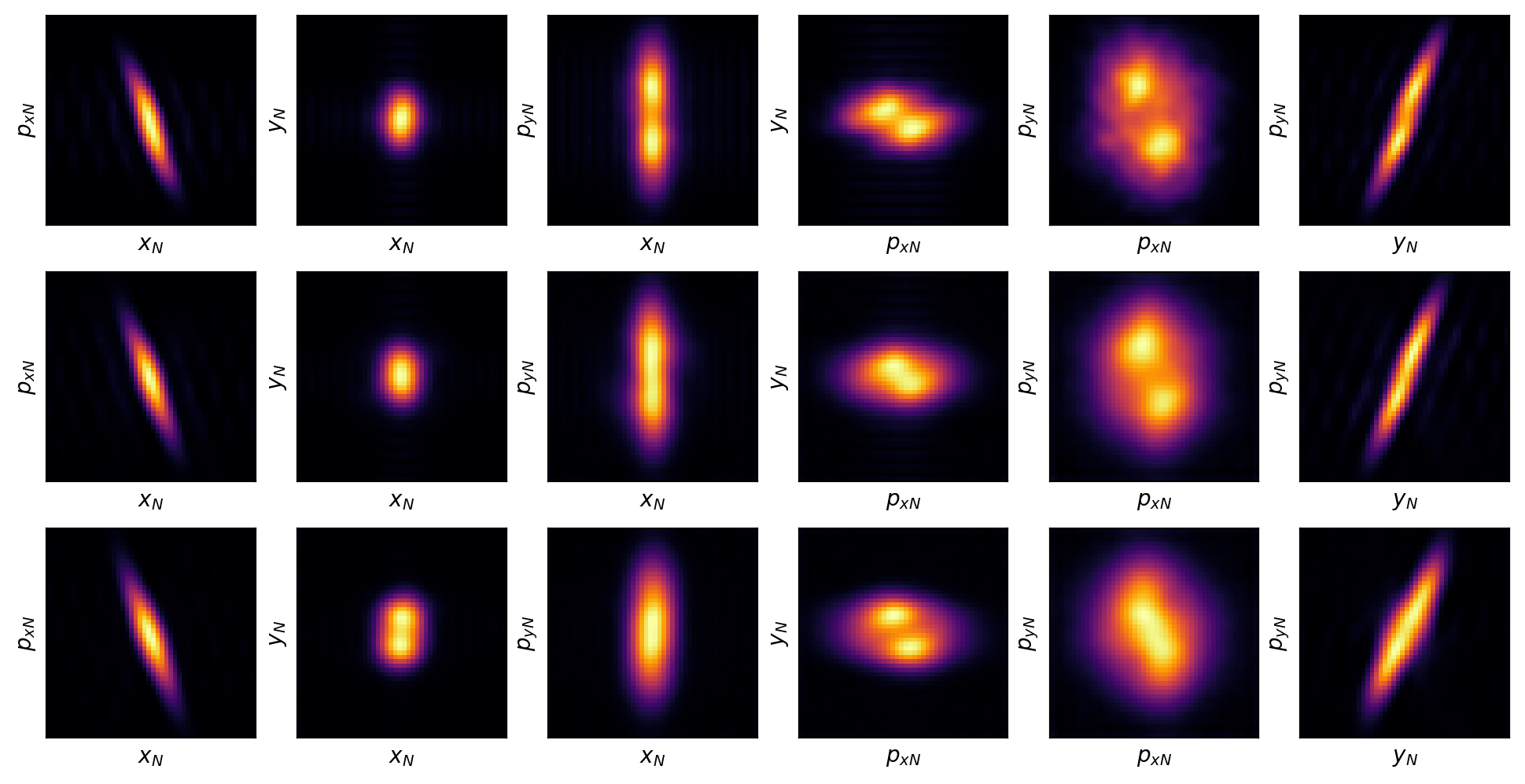} \\
Case C: quadrupole errors = (3.89\%,  -11.9\%, 8.89\%, -4.81\%, 14.0\%) \\
\includegraphics[width=.77\textwidth]{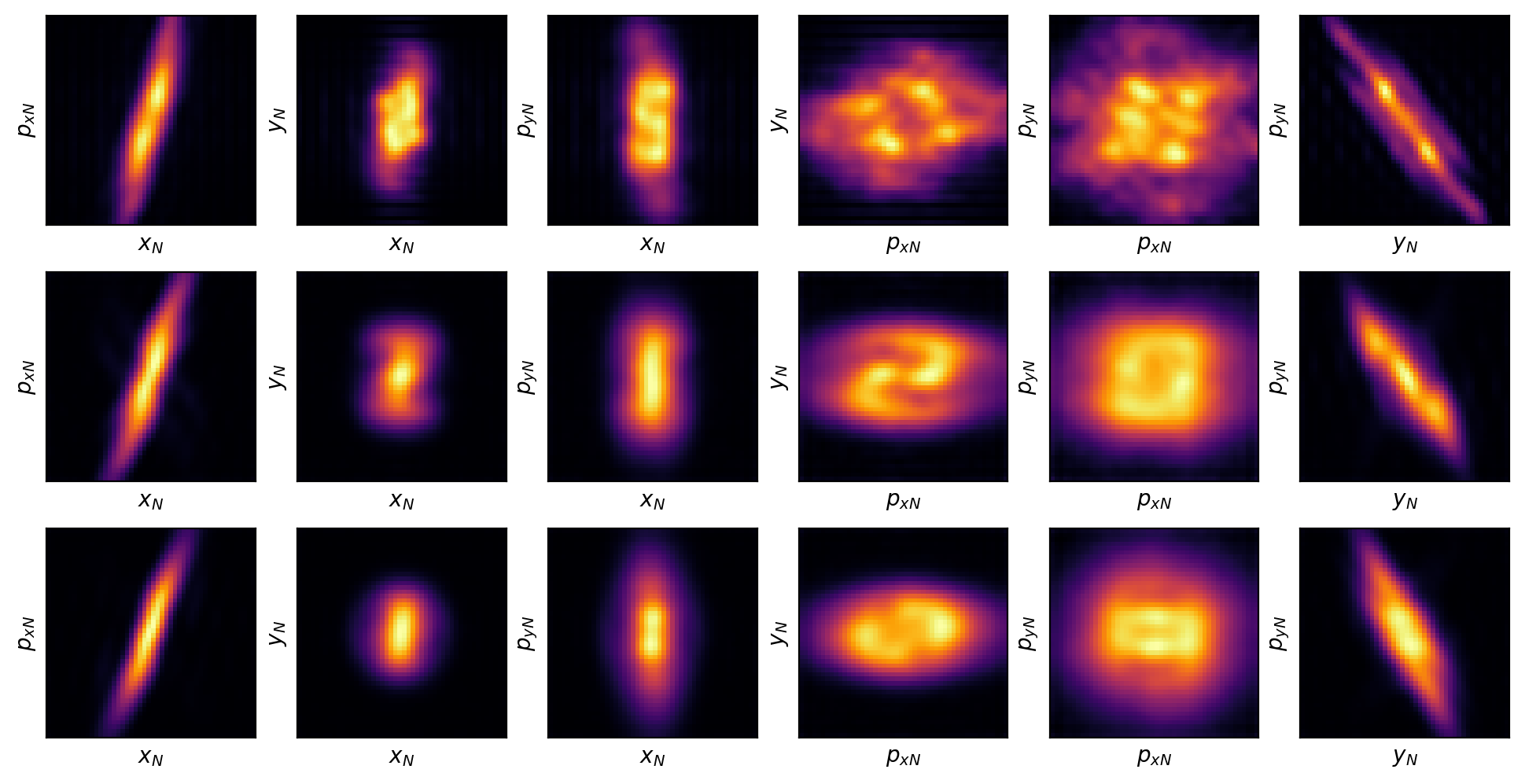}
\caption{Examples of phase space reconstructions from the phase space
decoder.  The cases are taken from the test data, and correspond to the 
phase space projections shown in figure 
\ref{fig:phasespaceprojectionsexamples}.  The top row in each case
shows the ground truth.  The second row shows the reconstruction from 
the sinogram that would be observed without quadrupole errors, using the
basic latent space from the sinogram autoencoder.  The third row shows
the reconstruction from the  sinogram with quadrupole errors, using the
extended sinogram encoder.
\label{fig:phasespacereconstructionsexamples}}
\end{figure}

Some examples showing the performance of the phase space decoder are
shown in figure \ref{fig:phasespacereconstructionsexamples}.  Three 
cases from the test data are shown: the top row in each case shows the 
ground truth (the phase space distribution used in simulation to
construct the sinograms in the training data).  The second row shows the
reconstruction of the phase space distribution from the sinogram
without errors, using the basic latent space from the sinogram 
autoencoder.  The third row shows the reconstruction of the phase space 
distribution from the sinogram with quadrupole errors, using the 
extended encoder.   The images in the second row illustrate how
well the machine learning techniques presented here can reproduce the
phase space distribution in the case of measurements without quadrupole
errors. The images in the third row illustrate the performance of the machine
learning techniques in the case of measurements in the presence of
quadrupole errors.  Although the reconstructed phase space distributions
do not reproduce the full level of detail in the `true' phase space 
distribution, the accuracy of the reconstruction in both cases (with or
without quadrupole errors) is still good, even in 
the presence of quadrupole strength errors that significantly affect the 
sinograms (see figure~\ref{fig:sinogramquaderrorsexample}).

\section{Experimental demonstration}
\label{sec:experimentaldemonstration}

To demonstrate the application of the extended latent space technique 
for phase space tomography, we present some experimental results from 
CLARA.  The data were collected in 2022, and a previous analysis was
reported in \cite{PhysRevAccelBeams.25.122803}.  The previous analysis
used machine learning, but did not account for possible errors on the
quadrupole strengths: the ML technique was based on a single neural
network taking the DCT of a sinogram as input, and providing the DCT of
the four-dimensional phase space distribution as output.  The parameters
(number of steps in the quadrupole scan, phase advances etc.) used for the
examples shown in sections \ref{sec:machinelearningprinciples} and 
\ref{sec:machinelearningmodel} in the current paper were chosen to allow
the trained neural networks to be applied directly to the data from 2022.
The data collected in 2022 included a case with bunch charge \SI{10}{pC}
and a case with bunch charge \SI{100}{pC}.  Here, we apply the new
technique to the case with bunch charge \SI{100}{pC}: in the previous
analysis \cite{PhysRevAccelBeams.25.122803}, it proved more difficult to
achieve good results with this data set, than for the case of a beam with
lower (\SI{10}{pC}) bunch charge.

\begin{figure}[htbp]
\centering
\includegraphics[width=\textwidth,trim={20mm 30mm 85mm 10mm},clip]{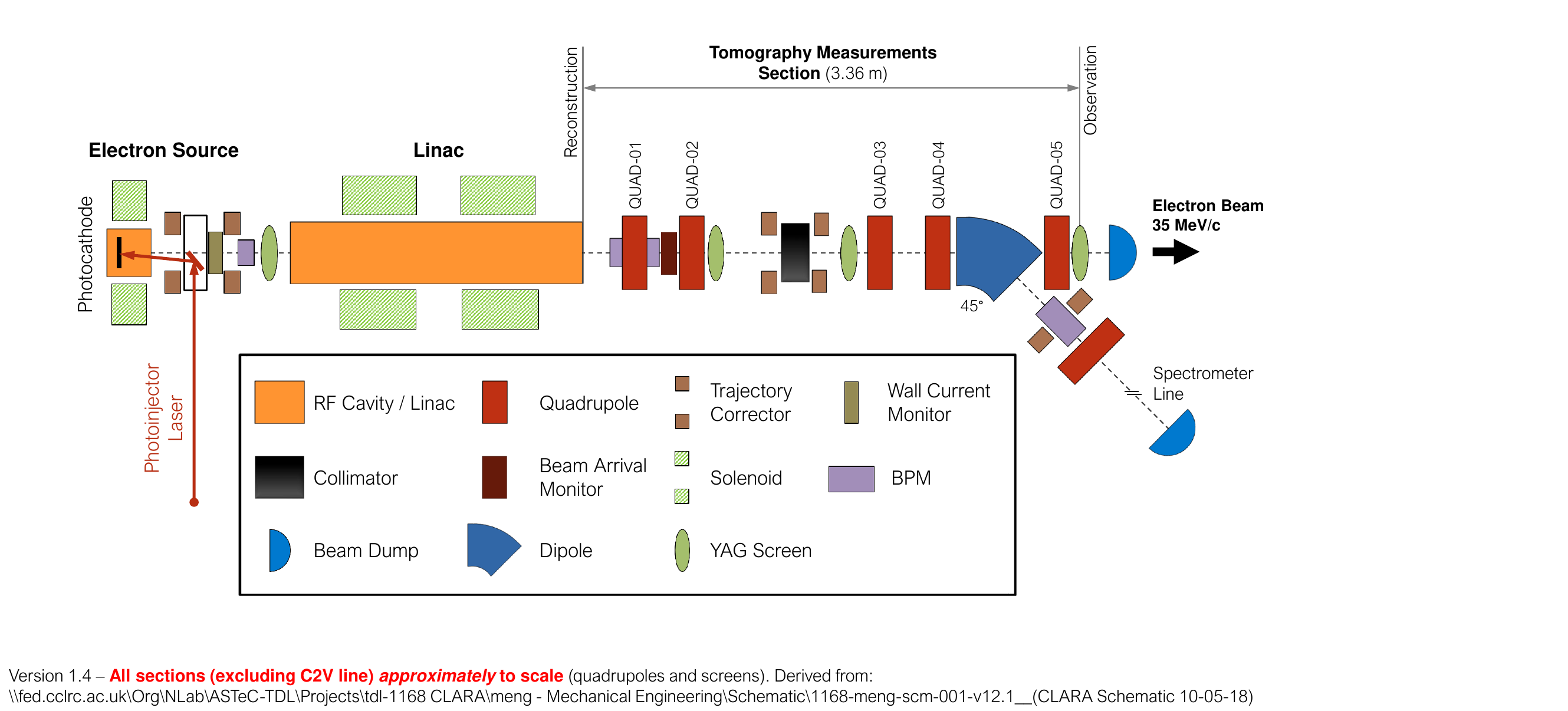}
\caption{Schematic of the CLARA accelerator test facility (not to scale).  The section
of beamline used for the tomography measurements is from the end of the linac
(the reconstruction point) to the third diagnostics YAG screen following the linac
(the observation point).  The five quadrupoles used in the quadrupole scan are
shown as red boxes (labelled QUAD-01, 02, 03, 04 and 05).
\label{fig:claraschematic}}
\end{figure}

A schematic of CLARA including the section of beamline used for the tomography
measurements is shown in figure~\ref{fig:claraschematic}.  For the
measurements, the machine was set up in a standard configuration with beam
energy \SI{35}{MeV} and bunch charge \SI{100}{pC}.  The reconstruction
and observation points are the end of the linac and the third diagnostics (YAG)
screen following the linac, respectively.  Five quadrupoles, labelled QUAD-01,
QUAD-02, QUAD-03, QUAD-04 and QUAD-05 in figure~\ref{fig:claraschematic}
were used in the quadrupole scan.

\begin{figure}[htbp]
\centering
\includegraphics[width=.95\textwidth]{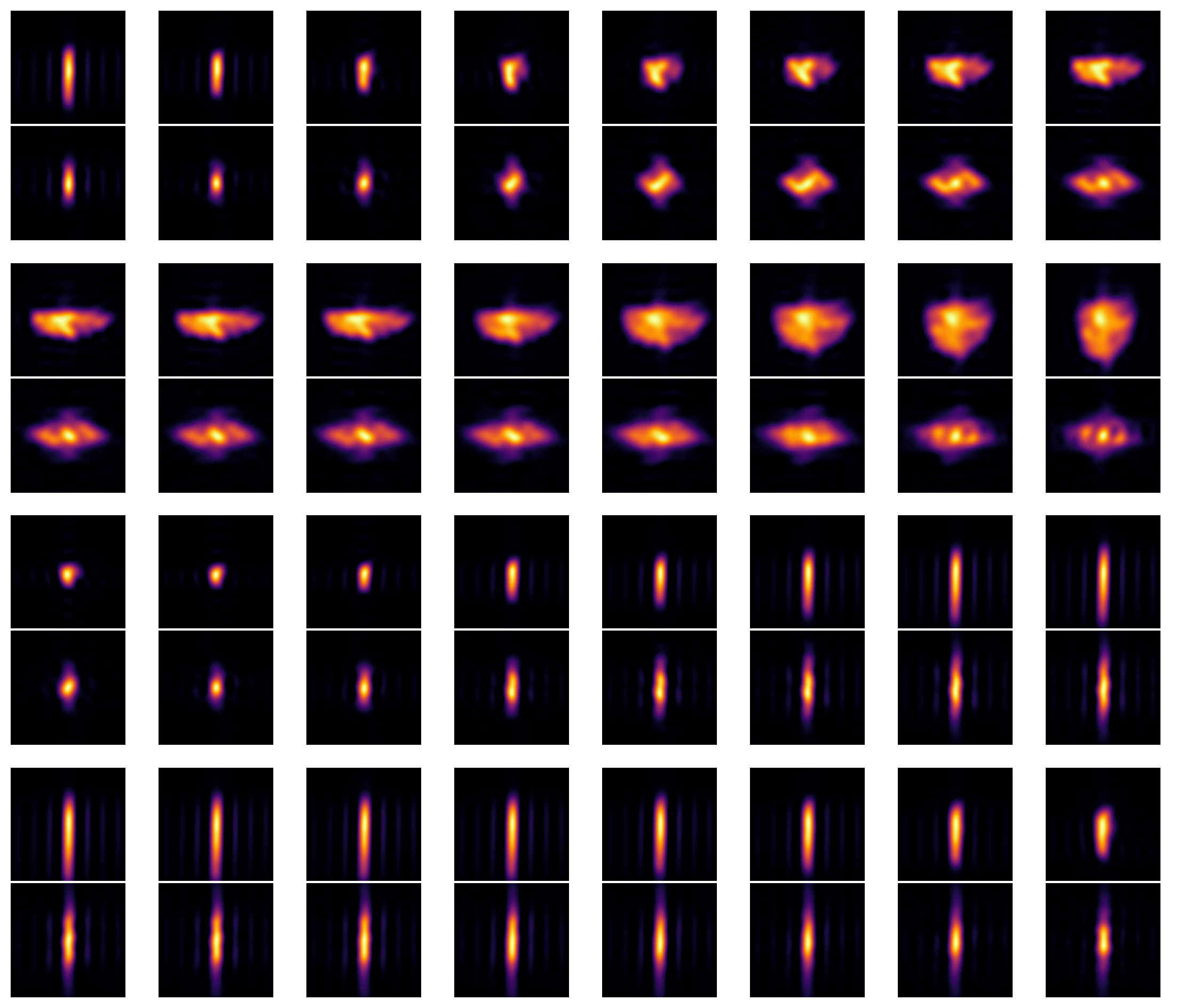}
\caption{Results from the sinogram autoencoder, for experimental data 
from CLARA. The top image in each pair shows the observed screen image, 
and the lower image shows the output of the autoencoder.  Note that the 
autoencoder takes no account of magnet errors, so if magnet errors are 
present during the observations a close agreement between the
autoencoder input and output images is not necessarily expected.
\label{fig:sinogramautoencoderexample100pC}}
\end{figure}

The screen images from the quadrupole scan are shown in figure 
\ref{fig:sinogramautoencoderexample100pC}, together with the output of
the sinogram autoencoder.  Note that the autoencoder does not take into 
account the possibility of errors on the quadrupole magnets.  Although 
the autoencoder reproduces reasonably well the overall variation in the
beam size over the course of the quadrupole scan, there are significant
differences in individual images: this may indicate that experimental
data were collected under conditions that do not accurately match the
model used to construct the training data for the autoencoder.  In other
words, differences between the observed images and the images from the
autoencoder may be the result of errors on components in the accelerator 
that are not taken into account in the simulation used to generate the
training data.

\begin{figure}[htbp]
\centering
\includegraphics[width=.95\textwidth]{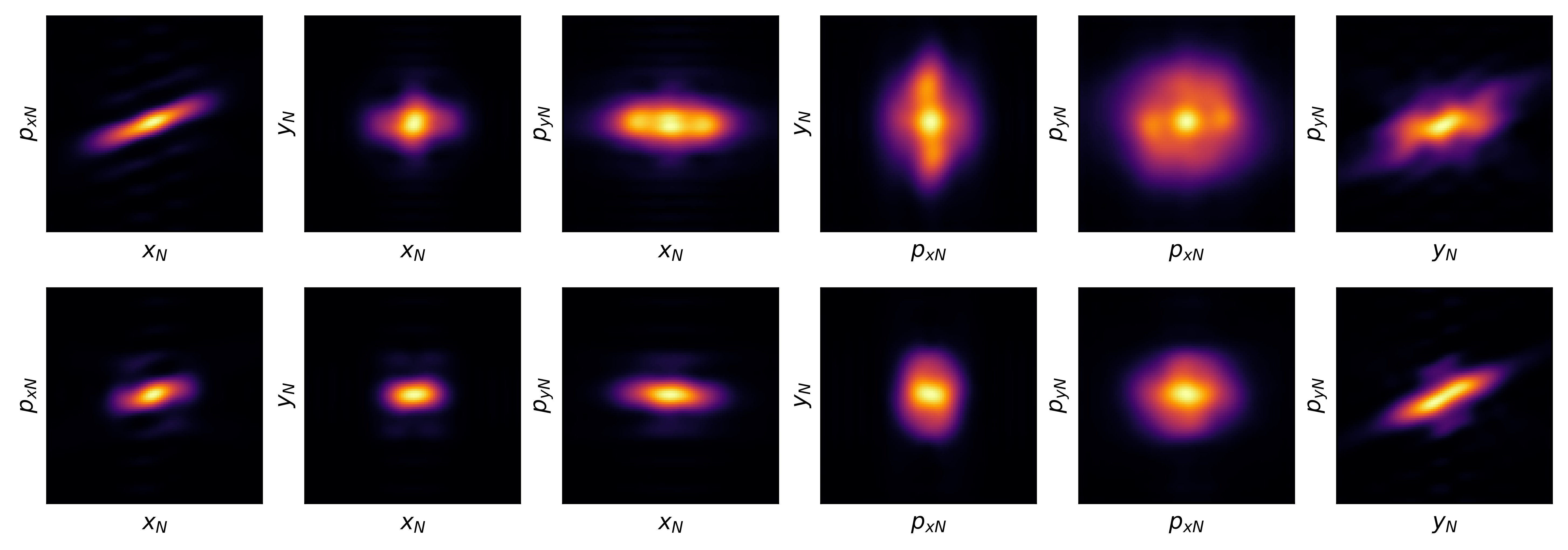}
\caption{Two-dimensional projections from the four-dimensional phase
space beam distribution reconstructed from quadrupole scan measurements 
in CLARA. The top row shows projections of the phase space distribution
found using the sinogram autoencoder and phase space decoder,
i.e.~without accounting for quadrupole strength errors.  The bottom row
shows projections from the phase space distribution found using the
extended encoder and phase space decoder, i.e.~taking into account the
effects of quadrupole strength errors.  Projections are onto different
pairs of normalised phase space variables, with axes range
$\pm$2\,mm/$\sqrt{\mathrm{m}}$.
\label{fig:clarameasuredphasespace100pC}}
\end{figure}

The phase space reconstructed from the measured sinogram in CLARA is
shown in figure \ref{fig:clarameasuredphasespace100pC}.  The top row in
this figure shows projections (in normalised co-ordinates) of the phase 
space distribution found using the sinogram autoencoder and phase space
decoder, i.e.~without accounting for strength errors on the quadrupole 
magnets.  The bottom row shows corresponding projections from the phase
space distribution found using the extended encoder and phase space 
decoder, i.e.~taking into account the effects of quadrupole strength 
errors. The projections in the top row of
figure~\ref{fig:clarameasuredphasespace100pC} can be compared with
those of the  previous analysis, shown in figure 12(b) in
\cite{PhysRevAccelBeams.25.122803}. Despite some differences, there
is general agreement in the shape of each projection.

For the two distributions in
figure~\ref{fig:clarameasuredphasespace100pC}, we observe that although
the shapes of the distributions are similar, the distribution reconstructed
using the extended encoder (taking quadrupole errors into account) appears
to occupy a smaller volume in phase space than the distribution reconstructed
using the sinogram autoencoder. For a Gaussian distribution, the emittance
provides a convenient measure of the phase space volume occupied by a
bunch of particles.  To estimate the emittances, each of the distributions
shown in figure~\ref{fig:clarameasuredphasespace100pC} is fitted with a
Gaussian of the form:
\begin{equation}
\hat{\rho} = \hat{\rho}_0 \exp\left(-\frac{1}{2}\mathbf{x}_N^\mathrm{T}\Sigma^{-1}\mathbf{x}_N \right),
\end{equation}
where $\Sigma_{ij} = \langle x_{N,i} x_{N_,j} \rangle$ is the covariance
matrix for the four-dimensional Gaussian in normalised phase space
(with phase space variables $x_{N,i}$), and $\hat{\rho}_0$ is the peak
density. The emittances can then be found from the covariance matrix
using:
\begin{equation}
\mathrm{eigenvalues}(\Sigma S) = \pm i \epsilon_k,
\end{equation}
where $\epsilon_k$ are the emittances (with $k = 1, 2$ for each of the two
normal modes in the transverse phase space), and the antisymmetric unit
matrix $S$ is defined by:
\begin{equation}
S = \left( \begin{array}{cccc}
 0 &  1 &  0 &  0 \\
-1 &  0 &  0 &  0 \\
 0 &  0 &  0 &  1 \\
 0 &  0 & -1 &  0
\end{array} \right).
\end{equation}
The emittances thus found from the phase space distributions
shown in figure~\ref{fig:clarameasuredphasespace100pC} are given in
table~\ref{tab:emittances}.  Taking errors on the quadrupoles into account
results in values for the beam emittances that are about a factor of 2
smaller than the values found without accounting for quadrupole errors. 

\begin{table}[htb]
\caption{Normalised emittances of the four-dimensional transverse phase
space distributions found using the sinogram autoencoder (not accounting
for quadrupole errors) and the extended autoencoder (accounting for
quadrupole errors), for experimental data in CLARA (with \SI{100}{pC}
bunch charge).  The relativistic factor $\gamma = 69.5$ corresponds
to a nominal beam energy of \SI{35.0}{MeV}.  
\label{tab:emittances}}
\begin{center}
\begin{tabular}{lcc}
\hline
 & $\gamma\epsilon_\mathrm{I}$ & $\gamma\epsilon_\mathrm{II}$ \\
 \hline
 sinogram autoencoder & 8.0\,$\upmu$m & 17.7\,$\upmu$m \\
 extended encoder & 4.8\,$\upmu$m & 7.9\,$\upmu$m \\
 \hline
\end{tabular}
\end{center}
\end{table}


As well as taking quadrupole focusing errors into account in reconstructing
the phase space distribution, the extended encoder provides an indication
of the sizes of the errors on the first three quadrupoles used in the quadrupole
scan.  For the CLARA data with \SI{100}{pC} bunch charge, the errors from
the extended encoder are found to be (12.7$\pm$3.4)\%, (12.7$\pm$2.8)\%
and (0.4$\pm$2.6)\%. These errors are larger than would normally be
expected; but it is worth noting that the first two quadrupoles have the same
error values.  The focusing errors in the quadrupoles may be the result of an
error on the beam energy rather than an error in the field gradients (see the
comments in section \ref{sec:trainingdata}): the focusing strength of a
quadrupole depends both on the field gradient in the magnet and on the
energy of the beam. If, during collection of the experimental data, the beam
energy was 10\% below the energy assumed in the analysis, then the
analysis would indicate an error of 10\% on the gradient of a quadrupole in
which the gradient was actually equal to the nominal gradient.  In the present
case, therefore, it is possible that the beam energy was about 12\% below
the expected value, while the first two quadrupoles had field gradients close
to nominal.  The third quadrupole, however, would then have a field gradient
significantly lower than expected. Unfortunately, this possibility cannot be investigated
further, because of subsequent changes to the accelerator made as part of work
to upgrade the facility.  However, although we do not present results
in detail in this paper, an analysis (using the extended encoder) of the
data for \SI{10}{pC} bunch charge suggests that the first two 
quadrupoles were close to their nominal focusing strengths, while the
third quadrupole was weaker by roughly 10\%.  The results from the
\SI{10}{pC} and \SI{100}{pC} cases would be consistent if it is
assumed  that the beam energy was lower than the nominal value for
the \SI{100}{pC} measurements, close to nominal for the \SI{10}{pC}
measurements, and that the third quadrupole had a gradient error of 
about -10\% for both cases.

\begin{figure}[htbp]
\centering
\includegraphics[width=.80\textwidth]{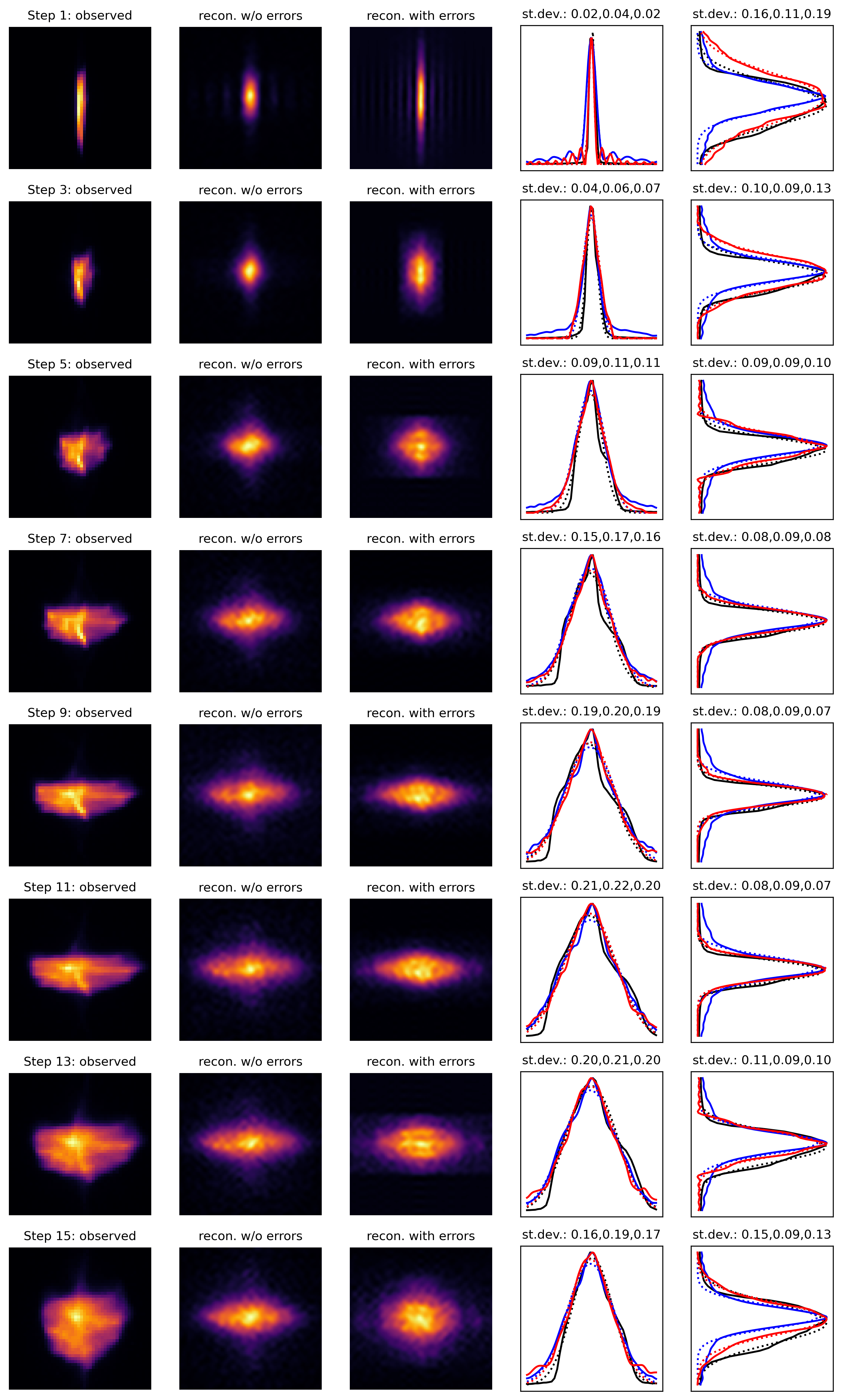}
\caption{Comparison between reconstructed and observed screen images
during the first half of a quadrupole scan in CLARA, with \SI{100}{pC}
bunch charge. The columns from left to right show: (1) original screen
image; (2) reconstructed image neglecting quadrupole errors;
(3) reconstructed image accounting for quadrupole errors; (4) and (5)
projections of each image onto the horizontal and vertical axes,
respectively. In columns 4 and 5 solid black, blue and red lines 
correspond to projections from images in columns 1, 2 and 3, 
respectively (dotted lines show Gaussian fits, with standard 
deviations shown above each plot).
\label{fig:experimentalsinogramx}}
\end{figure}

\begin{figure}[htbp]
\centering
\includegraphics[width=.80\textwidth]{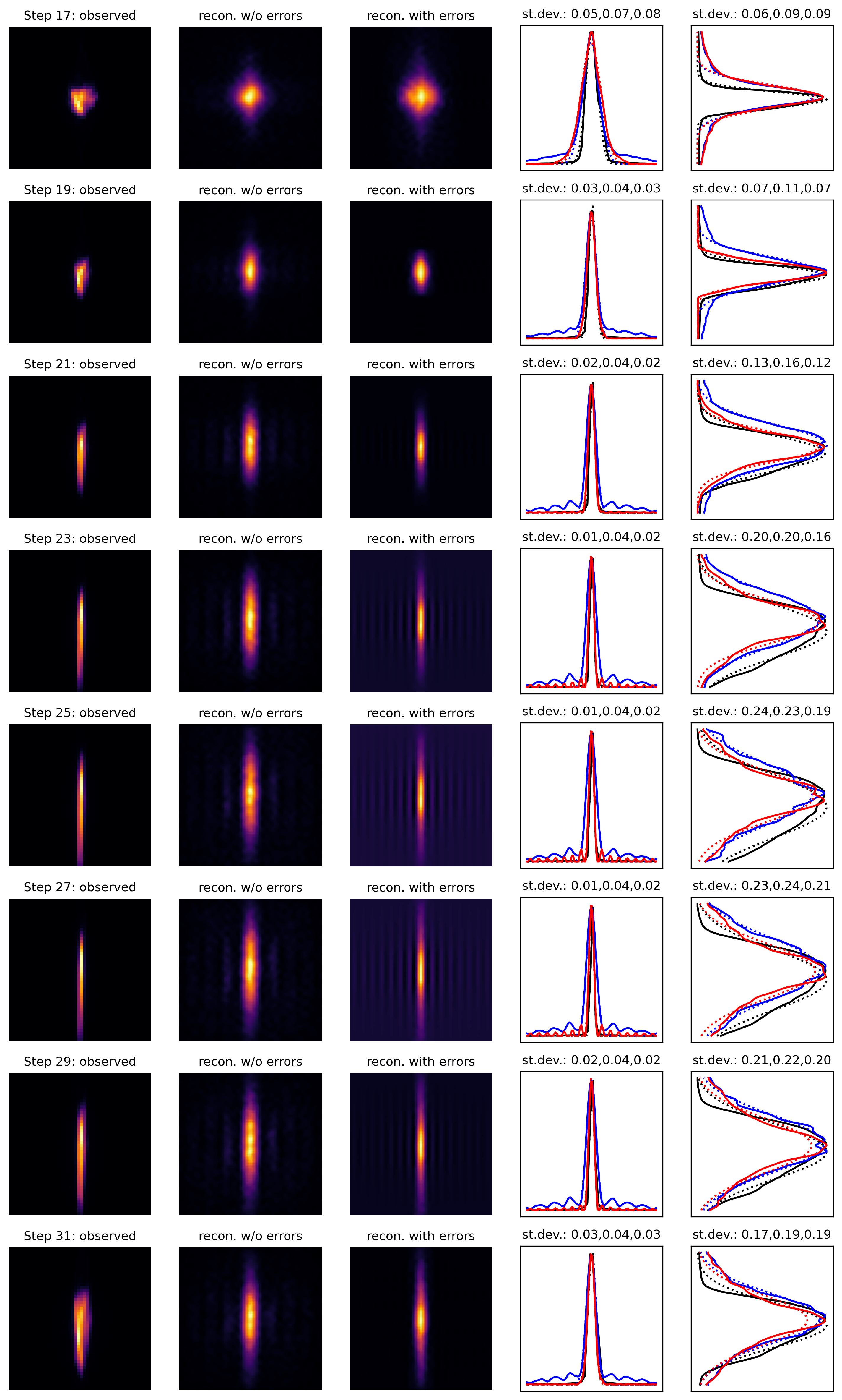}
\caption{Comparison between reconstructed and observed screen images
during the second half of a quadrupole scan in CLARA, with \SI{100}{pC}
bunch charge. The columns from left to right show: (1) original screen
image; (2) reconstructed image neglecting quadrupole errors;
(3) reconstructed image accounting for quadrupole errors; (4) and (5)
projections of each image onto the horizontal and vertical axes,
respectively. In columns 4 and 5 solid black, blue and red lines 
correspond to projections from images in columns 1, 2 and 3, 
respectively (dotted lines show Gaussian fits, with standard 
deviations shown above each plot).
\label{fig:experimentalsinogramy}}
\end{figure}

The reconstructed phase space distribution may be validated by
comparing the screen images predicted by a tracking simulation (using
the reconstructed phase space distribution and the estimated quadrupole 
strength errors) with the screen images observed during the 
measurements. Results of such a comparison are shown in
figures~\ref{fig:experimentalsinogramx} (for the first half of the
quadrupole scan) and \ref{fig:experimentalsinogramy} (for the second
half of the scan). Note that we show images from alternate steps in the
quadrupole scan, rather than all 32 images. In these figures, the first
three columns from left to right show: (1) the observed screen image;
(2) the image reconstructed from the phase space decoder applied to
the latent space from the sinogram autoencoder (i.e.~neglecting
quadrupole errors); (3) the image reconstructed from the phase space
decoder applied to the extended latent space (i.e.~accounting for
quadrupole errors). The reconstructed images are from particle tracking
simulations using the predicted phase space distributions, with quadrupole
errors included for the images in the third column. The final two columns in
figures~\ref{fig:experimentalsinogramx}
and \ref{fig:experimentalsinogramy} show projections of each image onto 
the horizontal and vertical axes, respectively. The solid black, blue and red
lines correspond (respectively) to projections from the observed image, the
reconstructed image neglecting quadrupole errors, and the reconstructed
image accounting for quadrupole errors. Dotted lines show Gaussian fits,
with standard deviations (as a fraction of the width of the image) shown
in the text above each individual plot.

Although the machine learning methods that we have applied in the present
analysis do not reveal the same level of detail in the phase space
distribution as in the earlier work \cite{PhysRevAccelBeams.25.122803},
we emphasise that a key
feature of the technique presented here is the use of a low-dimensional
latent space to allow the inclusion of parameters describing errors on 
accelerator components: some loss of detail is to be expected from 
reducing information about the phase space distribution to a set of just 
40 parameters in the latent space.

\begin{figure}[htbp]
\centering
\includegraphics[width=.95\textwidth]{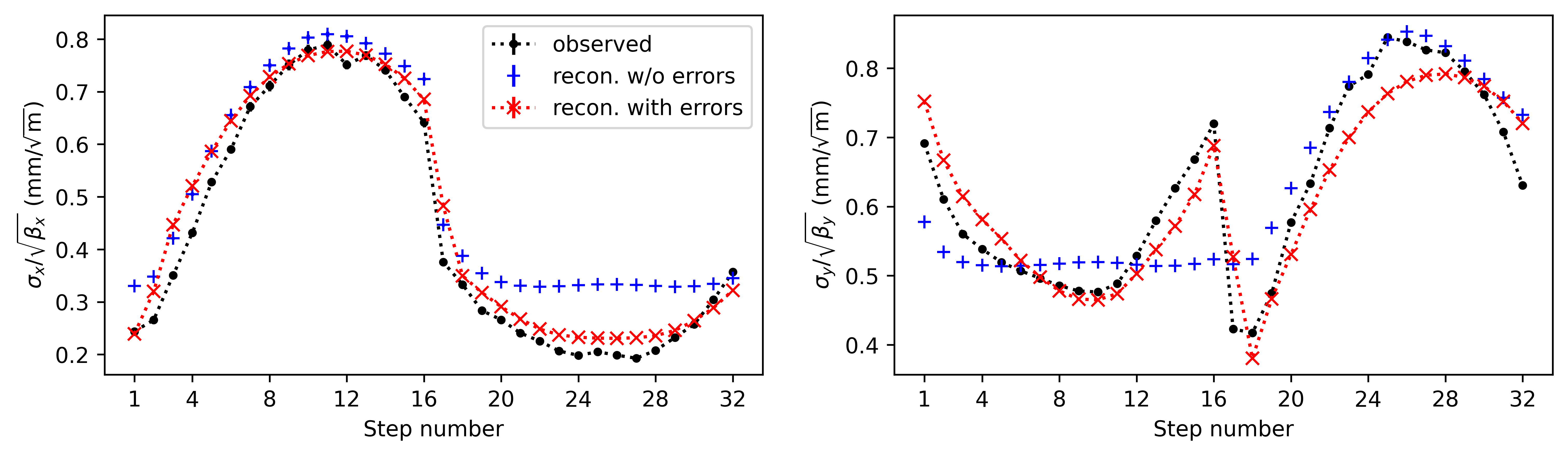}
\caption{Variation in beam sizes (scaled by the square root of the beta
function at the observation point) over the course of a quadrupole scan
measurement on CLARA, with \SI{100}{pC} bunch charge.  The left-hand
plot shows the horizontal beam size, and the right-hand plot shows the
vertical beam size.  The beam sizes are obtained as the widths of 
Gaussian fits to the respective images projected onto the horizontal or
vertical axes.  Error bars (which are small) indicate the uncertainty
on the fit.
\label{fig:beamsizevariation}}
\end{figure}

A simplified comparison between the observed screen images and the
reconstructed images (neglecting or taking account of quadrupole
errors) is presented in figure~\ref{fig:beamsizevariation}. The plots show 
the variation in horizontal and vertical beam sizes at the observation
point over the course of the quadrupole scan, scaled by the square root
of the beta function at the observation point (so that if the phase
space distribution was correctly described by the lattice functions, the 
scaled beam size would be consant). The beam sizes are
obtained from the widths of Gaussian fits to the projections of the
respective screen images onto the horizontal and vertical axes.
Although there are some differences between the observed and the
reconstructed beam sizes, the overall level of agreement appears
reasonable.  The results in figure~\ref{fig:beamsizevariation} can be
compared with those shown in figure 14 in
\cite{PhysRevAccelBeams.25.122803}.  We see in particular that taking
account of errors on the quadrupole magnets leads to significantly
better agreement in the vertical beam size in steps 12 to 16 of the
scan.  Note, however, that the reconstruction without errors shown in
figure~\ref{fig:beamsizevariation} appears to given worse agreement
(for some of the steps in the quadrupole scan) with the measured beam
sizes than the results shown in figure 14 in
\cite{PhysRevAccelBeams.25.122803}.
Although we have not investigated the reasons in detail, it is possible
that the previous analysis gave better agreement because of the
way in which the neural network was designed to produce the
phase space distribution directly, without an intermediate latent space.
Without the dimensionality reduction resulting from the use of a latent
space, the neural network would have greater flexibility to `fit' the phase
space distribution to the observed data even without taking advantage of
the additional degrees of freedom allowed by explicit inclusion of quadrupole
errors. 

\section{Conclusions}
\label{sec:conclusions}


An increasing number of studies are demonstrating the value of machine
learning techniques for accelerator diagnostics.  Even where alternative
methods for data analysis exist, machine learning can significantly reduce
the time needed for processing complex data; but machine learning
techniques can also make it possible to extend diagnostics tools beyond
previous capabilities.  For phase space tomography, there are well-established
conventional algorithms that can be used, even in the case of
higher-dimensional phase space reconstructions; but it is difficult, using these
algorithms, to take account of machine errors. Given the complexity of
accelerator systems, some variation of parameters (including magnet
field strengths, RF field amplitudes and phases etc.) from their specified
values is always possible; such variations can have a significant impact
on beam behaviour, and it is therefore important to allow for the
possibility of errors on components in an accelerator during any beam
measurement.  Machine learning techniques offer the possibility of
taking account of machine errors in phase space tomography. Furthermore,
once trained, a machine learning model can be used to reconstruct the phase space
distribution in two degrees of freedom from a set of beam images much
more quickly than by using conventional (non-ML) methods.  

Taking advantage of the capabilities offered by machine learning can
present some challenges.  For example, for a technique involving supervised
learning, there is the problem of generating or collecting sufficient training
data. This often means depending on simulations, since for many problems
it will not be possible to make a direct observation of the required output:
this is the case, for example, in beam phase space tomography.  Even
where it is possible, in principle, to use experimental observations for training
a machine learning model, the time required to collect sufficiently large data
sets can make such an approach impractical.

A further challenge in the use of machine learning in accelerator diagnostics
is validating the results. In the case of phase space tomography, it is
possible to assess the accuracy of the reconstructed phase space distribution
by simulating the measurements made in collecting the data, i.e.~by
reproducing the screen images.  Ultimately, though, the most important
test of any diagnostics measurement is whether the results are of value
in tuning and operating the accelerator, which usually means being able to
make accurate predictions, based on the diagnostics results, of the
response of the accelerator to changes in the settings of different
components.

The dependence in many cases on simulations for the application of
machine learning techniques raises the importance of an accurate
computer model of the accelerator.  In this respect, a technique such
as that presented in this paper (aiming to provide information about the
accelerator components used in a measurement as well as the main
goals of the measurement) can be of significant value, even if the
information obtained about the measurement components is subject to
some uncertainty.  This is illustrated in the case of the CLARA data
discussed in section~\ref{sec:experimentaldemonstration}, where
taking some account of errors in accelerator components
improves the match between the beam size variation observed
during a quadrupole scan, and the variation predicted by the tomography
results.  Even though there remains some uncertainty regarding the
(apparent) errors on the quadrupole focusing strengths, the analysis
does provide some information that can be useful for further investigation
and correction of possible errors.

Finally, as we observed in section~\ref{sec:machinelearningprinciples},
there are many different ways in which machine learning tools can
be applied for phase space tomography.  Understanding the most
appropriate technique for a given problem, and optimizing the
implementation (including the measurement parameters, machine
learning architecture, and training data characteristics) is a
formidable task, and will require further work.

\acknowledgments

We would like to thank our colleagues in STFC/ASTeC for help with the
collection of experimental data, in particular Peter Williams, Boris
Militsyn and Matthew King.  This work was supported by STFC through
a grant to the Cockcroft Institute.



\bibliographystyle{JHEP}
\bibliography{biblio.bib}







\end{document}